\documentclass[iop]{emulateapj}
\usepackage{natbib, bm, amsmath}
\shorttitle{Toward a New Kind of Asteroseismic Grid Fitting}
\shortauthors{Gruberbauer, Guenther \& Kallinger}

\begin{document}

\def \uHz {\,$\mu$Hz}

\title{Toward a New Kind of Asteroseismic Grid Fitting}

\author{M. Gruberbauer and D.~B. Guenther}
\affil{Institute for Computational Astrophysics, Department of Astronomy and Physics, Saint Mary's University, B3H 3C3 Halifax, Canada}
\author{T. Kallinger}
\affil{Instituut voor Sterrenkunde, K.U. Leuven, Celestijnenlaan 200D, 3001 Leuven, Belgium}

\affil{Institute for Astronomy, University of Vienna, T\"urkenschanzstrasse 17, 1180 Vienna, Austria}

\begin{abstract}
Recent developments in instrumentation (e.g., in particular the {\it Kepler} and {\it CoRoT} satellites) provide a new opportunity to improve the models of stellar pulsations. Surface layers, rotation, and magnetic fields imprint erratic frequency shifts, trends, and other non-random behavior in the frequency spectra. As our observational uncertainties become smaller, these are increasingly important and difficult to deal with using standard fitting techniques. To improve the models, new ways to compare their predictions with observations need to be conceived. In this paper we present a completely probabilistic (Bayesian) approach to asteroseismic model fitting. It allows for varying degrees of prior mode identification, corrections for the discrete nature of the grid, and most importantly implements a treatment of systematic errors, such as the ``surface effects." It removes the need to apply semi-empirical corrections to the observations prior to fitting them to the models and results in a consistent set of probabilities with which the model physics can be probed and compared. As an example, we show a detailed asteroseismic analysis of the Sun. We find a most probable solar age, including a $35 \pm 5$ million year pre-main sequence phase, of 4.591 billion years, and initial element mass fractions of $X_0 = 0.72$, $Y_0 = 0.264$, $Z_0 = 0.016$, consistent with recent asteroseismic and non-asteroseismic studies. 
\end{abstract}


\section{Introduction}
The success of recent space missions {\it CoRoT} and {\it Kepler}, designed for the discovery of exoplanets and the analysis of stellar pulsation, have produced a large number of high-quality light curves \citep{chaplin2010}. With these data sets, obtained over long time bases of several months, we are able to detect variability with semi-amplitudes down to a few parts per million. These observations have now firmly established the existence of solar-type pulsation in a large number of solar-like and red-giant stars. Moreover, observations of an unprecedented number of $\delta$\,Scuti stars and other types of pulsators have also revealed rich mode spectra.

These data are now causing a paradigm shift for many topics in stellar astrophysics. In particular, the determination of fundamental stellar parameters, and any inferences regarding the physics of stellar interiors, have for a long time been restricted to testing theoretical models using classic observables such as photometric indices or spectroscopic data. Even though these methods have become more advanced, for instance by applying complex Bayesian methods to determine stellar ages \citep{pont2004, jorgensen2005} and to evaluate competing models \citep{takeda2007, bazot08}, the value of additional information provided by pulsation modes is tremendous, as they directly probe the whole star.
Already, the asteroseismic community is successful in extracting general characteristics of the mode spectra for many different types of stars \citep[e.g.,][]{mathur2010, kallinger2010a} and also in devising promising tools for a comparative interpretation of the observations \citep[e.g.,][]{bedding2010a}. 
Average mode parameters, such as the large and small frequency separations, and the frequency of maximum power, have been shown to  successfully constrain stellar parameters although certain correlations remain as a source for uncertainty \cite[see, e.g.,][]{kallinger2010c,huber2011, gai2011}.
These have been incorporated into the current advanced probabilistic pipelines to investigate stellar model grids \citep{quirion2010} and already been applied to recent observations \citep{metcalfe2010}. The next step to improving our knowledge about stellar interiors is to analyze individual pulsation modes in an equally rigorous way, to see where our models agree or disagree.

In the past, $\chi^2$-minimization techniques \citep{guenther2004}, or equivalent Bayesian analyses \citep[e.g.,][]{kallinger2010b}, have been introduced to find the pulsation model that most closely reproduce the observed frequencies within a large and dense grid of models. The Bayesian analysis, in this context, only provides an additional framework for constraining solutions to models that match our prior knowledge about the stars' fundamental parameters. Due to the rich information provided by the pulsation frequencies, these approaches should be successful in many cases, which is why they are being applied also to the most recent {\it Kepler} data sets. For instance, \cite{metcalfe2010} test various approaches from different modelers with different methods that actually use the individual frequencies. However, there are currently (at least) three major problems when applying these techniques. 
\newline
\newline
{\it Stellar rotation}, at all but the slowest rotation speeds, has been shown to produce rotational splittings which are incompatible with the traditional linear approximations. It even perturbs the values of the axisymmetric $(m=0)$ frequencies \citep[e.g., see][and references therein]{deupree2010}. In order to correctly take this into account, the rotation speed as a function of stellar depth needs to be known, and extensive computations would be necessary to do these effects justice. Given the large variety of possible rotation profile characteristics, this would greatly expand the dimensionality and size of the pulsation model grid. This implies currently insurmountable computational expenses for the types and sizes of grids that are necessary for a comprehensive asteroseismic analysis of many stars. 

\begin{figure}[t!]
   \centering
   \includegraphics[width=\columnwidth]{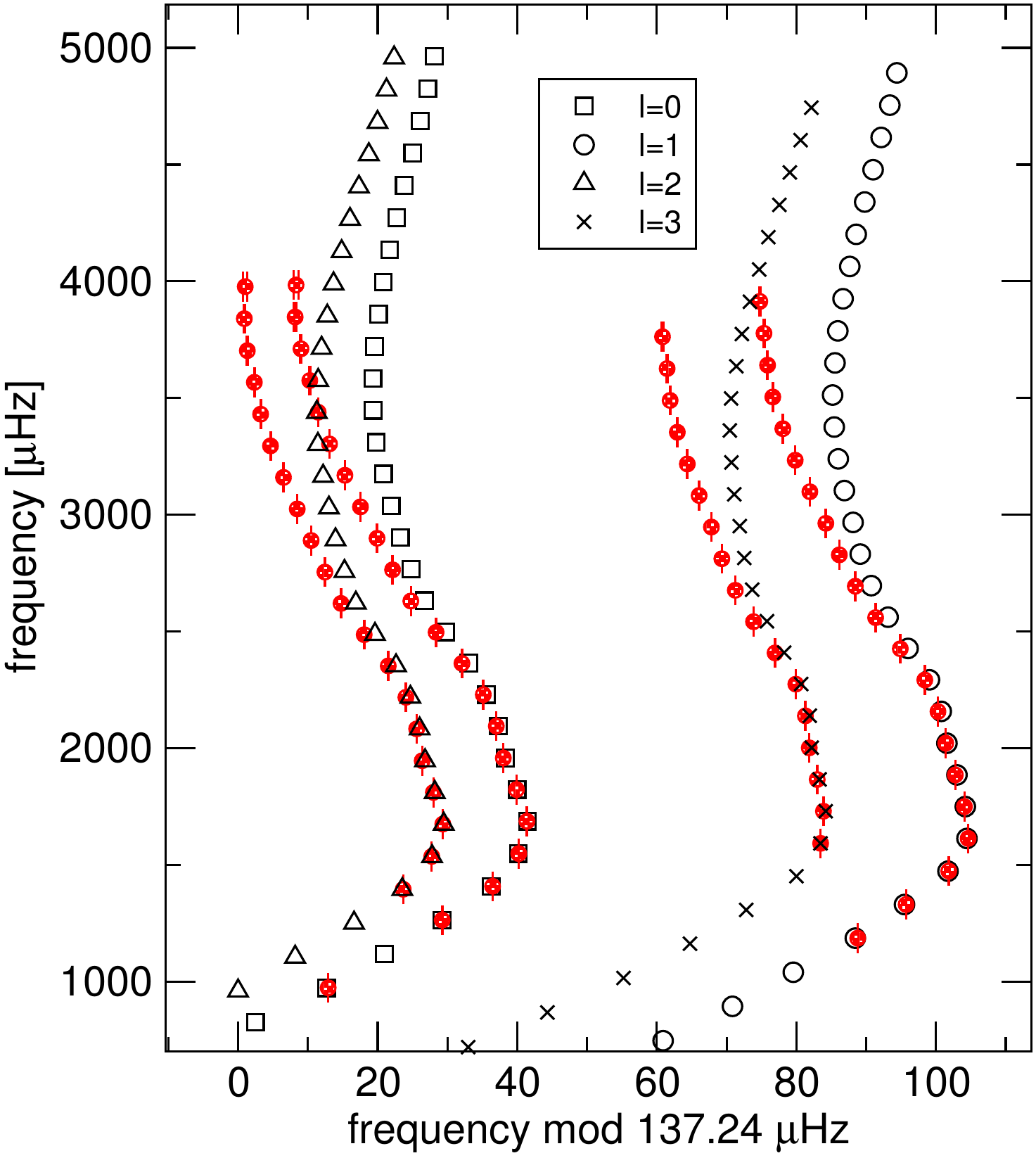} 
   \caption{Echelle diagram of solar {\it p} modes taken from \cite{broomhall2009} (filled circles) and an appropriate solar model
   constructed using YREC. The higher order model frequencies are increasingly deviating from the observations due to deficiencies
   in modeling the upper stellar layers. The systematic errors of the models are much bigger than the random observational uncertainties.}
   \label{fig:solarsurf}
\end{figure}

For stars with a convective envelope, model frequencies at high radial orders differ from observations due to problems in modeling the outer layers (see Figure\,\ref{fig:solarsurf}). These so-called {\it surface effects} can be compensated by looking at ratios of frequency differences \citep{roxburgh2005}, or by ``correcting" the observed frequencies through calibration of the surface effects seen for the Sun as proposed by \cite{kjeldsen2008}. It is likely that the surface correction as calibrated for the Sun is not universally applicable, and evidence for this has been mounting \citep[e.g.,][]{bedding2010b}. Moreover, neglecting (or correcting for) the surface effects in the observed frequencies is only reasonable when studying properties of the star for which the outer layers are unimportant. However, if we want the theoretical models to more closely reflect reality, we need to include more and better physics to bring the computed frequencies closer to the (un-corrected) observed ones. 

Furthermore, the fact that  {\it static asteroseismic grids can only have a finite resolution} in parameter space is often neglected. If the error bars of the observed frequencies are small compared to the differences between calculated frequencies in adjacent grid points, the likelihood of having a model in the grid that corresponds to the best model one's code could deliver decreases rapidly. The problem of finding the ``true" model and the actual uncertainties with respect to the grid becomes apparent. Even grids with adaptive resolution have the same problem in principle, as the decision for further refining the resolution of a particular region in parameter space must always depend on a number of discrete grid points. This problem is much more severe if our aim is to calculate probabilities (or some summary statistics) to compare different model grids.

In this paper we present a new approach to asteroseismic model grid fitting. Our goal is to find a new way of putting our model physics to the test that can handle all of the aforementioned difficulties. Even restricted to models that are unable to produce all the details of the observations, we want to know which models are most ``correct" (i.e., consistent with appropriate fundamental parameters and physics), and how well the solution is constrained. We show how to quantitatively assess our model grids as a function of the observational uncertainties, the uncertainties of the calculated frequencies, and our general prior knowledge about the star and possible shortcomings of our models.


\section{Bayesian treatment of systematic errors}

\subsection{Basics of Bayesian inference}
\label{sec:bayesbasics}
Bayes' theorem, applied to the problem of inference, states that the probability of a particular hypothesis after obtaining new data (i.e., the posterior) is proportional to the probability of the hypothesis prior to obtaining the new data (i.e., the prior) times the likelihood of obtaining the new data, under the assumption that the hypothesis is true (i.e., the likelihood function). This approach to inference is derived from the product and sum rules of probability theory that have shown to be necessary and sufficient for consistent, quantitative logical reasoning\footnote{Strictly speaking, Bayes' theorem is only one result that derives from these rules. Consistent use of Bayes' theorem, in particular the assignment of the various terms in Equation\,(\ref{equ:bayes0}), also requires knowledge of its origin and consistent application of the product and sum rule. However, for the sake of brevity we will simply call our approach in this manuscript to be ``Bayesian" rather than ``based on probability theory as extended logic".}\citep[see][]{jaynes}. 

In this paper, we stay as close as possible to the general notation used in \cite{jaynes} or \cite{gregory2005}. We start with Bayes' theorem applied to the problem of comparing observations with the predictions of a model $M$. If the predictions of a model $M$ are governed by a set of $n$ parameters $\boldsymbol{\theta}=\left\{\theta_1, ..., \theta_{n}\right\}$, and we define the observations to be represented by the symbol $D$ (for data), it is commonly formulated by expressing the posterior probability
\begin{equation}
P(\boldsymbol{\theta} | M, D, I) = \frac{P\left(\boldsymbol{\theta} | M, I \right) P\left( D | \boldsymbol{\theta}, M, I \right )}{P\left(D | M, I \right)}.
\label{equ:bayes0}
\end{equation}
The symbol $I$ is equivalent to the prior information about the problem that is investigated. The first term in the numerator of Equation\,(\ref{equ:bayes0}) is the prior probability of a particular set of parameter values $\boldsymbol{\theta}$, given the model $M$ and our prior information $I$ about the problem. It is independent of any new data which are supposed to be analyzed. The second term in the numerator is called the likelihood. It gives the likelihood of obtaining the observed data under the assumption that the predictions of model $M$ are correct, given the particular choice of its parameter values $\boldsymbol{\theta}$. The denominator in Equation\,(\ref{equ:bayes0}) is called the global likelihood, or {\it evidence}, and is the sum (or integral) of the numerator over the whole parameter space of model $M$. It therefore acts as a normalization constant. Most importantly, if the prior probabilities are adequately normalized, it also represents the likelihood of obtaining the data given the {\it whole} model $M$, independent of the particular choice of $\boldsymbol{\theta}$. Thus, it can be used as a likelihood for comparisons among different alternative models.  

More details on the application of Bayes' theorem, in particular with respect to data analysis in astronomy, can be found in \cite{gregory2005}. 


\subsection{Systematic errors in the Bayesian framework}

One of the strengths of the Bayesian framework is that a parameter $\theta_{n}$, known to be necessary to describe a model $M$, can be marginalized by applying the sum rule. In the case of continuous parameters, the sum turns into an integral, and by integrating the full posterior over the parameter range of $\theta_{n}$, one obtains the marginal posterior 
\begin{equation}
P(\theta_1, ..., \theta_{n -1} | M, D, I) = \int P(\theta_1, ..., \theta_{n -1}, \theta_{n} | M, D, I) \, d\theta_{n}.
\label{equ:margin}
\end{equation}
The marginal posterior retains the overall effects of including parameter $\theta_{n}$ in the model, but is independent of any particular choice of its value. In other words, $\theta_{n}$ is ``removed" from the detailed analysis. This is similar to what is done for calculating the evidence in the denominator in Equation\,(\ref{equ:bayes0}). The only difference is that the evidence is the marginal likelihood over all parameters of the model, weighted by the prior.

The reason this is useful is that if the data and the model are known to show systematic differences, like shifts or trends, such ``systematic errors" can simply be encoded introducing additional parameters to the model $M$, so that $M$ is able to model these effects as well. By subsequently marginalizing over these ``fudge parameters", one is then able to perform a standard Bayesian analysis without any need for knowing the exact value of the systematic error(s). However, even though the exact value is unknown, the presence of the error is being considered in the evaluation of the posterior probabilities. Furthermore, an increasing number of ``fudge parameters" comes at a cost, because it potentially decreases the evidence for the model due to the increase in prior volume. As mentioned in Section\,\ref{sec:bayesbasics} the evidence is used as the value for the likelihood of obtaining the data in Bayesian model comparison. It is therefore possible to compare models with and without ``fudge parameters". Improved models that do not need them, but are able to explain the observations just as well, will be favored. 
\newline
\newline


\section{Toward a Bayesian solution to asteroseismic model fitting}
\subsection{Review and problems of the standard approach}
\label{sec:standardapp}

The general problem of asteroseismic model fitting is to match observed frequencies $f_{i, \rm o}$ to those calculated from models $f_{i, \rm m}$. If the $n_{\rm obs}$ observed frequencies have individual uncertainties $\sigma_{i, \rm o}$, and the model frequencies have random uncertainties $\sigma_{i, \rm m}$ then a $\chi^2$-statistic can be calculated according to
\begin{equation}
\chi^2 = \frac{1}{n_{\rm obs}}\sum_{i=1}^{n_{\rm obs}} \frac{\left(f_{i, \rm o} - f_{i, \rm m}\right)^2}{\sigma_{i, \rm o}^2 + \sigma_{i, \rm m}^2}.
\label{equ:chisqu}
\end{equation}
Searching a large grid of $N$ stellar models $M_j$ with fundamental parameters close to those estimated for the observed star will produce a minimum in $\chi^2$ (= best-fit model). In addition, uncertainties can be estimated from the change in $\chi^2$ as the distance in parameter space to the best fit increases. Calculated with adequate stellar evolution and pulsation codes, it should be possible to infer details about the stellar interior and to obtain precise fundamental parameters.

In order to consistently encode prior information about the fundamental parameters and other model properties, and to make use of all the additional advantages that come with the Bayesian approach (all of which will become clear in the next section), it is much easier to perform the model fitting using probabilities. Assuming that the {\it random} uncertainties $\sigma_{i, \rm o}$ and $\sigma_{i, \rm m}$ are compatible with Normal distributions, one can define
\begin{equation}
\sigma_{i}^2 = \sigma_{i, \rm o}^2 + \sigma_{i, \rm m}^2. 
\end{equation}
\\
This leads to the likelihood for observing the data (= the specific values of $f_{i, \rm o}$), given a single observed and calculated frequency 
\begin{equation}
P(f_{i, \rm o} | f_{i, {\rm o} \mapsto i, {\rm m}}, M_j, I) = \frac{1}{\sqrt{2\pi}\sigma_{i}}\exp{\left[-\frac{\left(f_{i, \rm o} - f_{i, \rm m}\right)^2}{2\sigma_{i}^2}\right]}.
\label{equ:singleprob}
\end{equation}
Here, $f_{i, {\rm o} \mapsto i, {\rm m}}$ stands for the proposition ``{\it The observed mode $f_{i, \rm o}$ corresponds to the calculated mode $f_{i, \rm m}$}."\footnote{Although the explicit notation seems clumsy at first glance, it is actually one of the major assets of the Bayesian approach. It visualizes exactly which propositions we are evaluating, and under which conditions the probabilities are calculated. Slightly different propositions or conditions can yield vastly different results. If the notation is explicit, there are no hidden variables or assumptions.}
Naturally, we want our models $M_j$ to reproduce all observed frequencies. Assuming that each observed frequency is a statistically independent datapoint, this leads to a product for the likelihood of obtaining all observed frequency values given that the model is correct
\begin{equation}
P(D | M_j, I) = \prod_{i=1}^{n_{\rm obs}} P(f_{i, \rm o} |  f_{i, {\rm o} \mapsto i, {\rm m}}, M_j, I).
\label{equ:probdens}
\end{equation}
Here, $D$ stands for complete set of observed frequencies and their uncertainties. This can then be incorporated in the usual framework for Bayesian inference. 

Alas, both of the mentioned, straightforward approaches above suffer from the following problems: 
\begin{enumerate}
\item The most appropriate model is not necessarily the one that minimizes Equation\,(\ref{equ:chisqu}) or maximizes Equation\,(\ref{equ:probdens}). There are many possible scenarios where this would be the case (e.g., due to surface effects, stellar activity, magnetic field effects, rotational effects). Straightforward application of the formalism above will then lead to wrong or nonsensical results in both best fit and derived uncertainties. Even worse, this would propagate into our assessment of the model physics that were used to produce the models.
\item In case of such systematic differences, we need to take into account that for each observed pulsation frequency, multiple model frequencies are possible candidates (not necessarily only the closest one). This is particularly problematic in cases were no prior mode identification is available.
\item As the observational uncertainties decrease, the contrast in $\chi^2$ (and even more so the contrast in probabilities) between different models increases. If the model that minimizes/maximizes Equation\,(\ref{equ:chisqu})/Equation\,(\ref{equ:probdens}) is not the correct model due to missing physics, this increase in fitting contrast is misleading and unwarranted.
\item In static grids the finite grid resolution increases the risk of missing the most adequate model that the code could produce. If there are systematic differences between even the most adequate mode and the observations, the ``contrast enhancement effect" will be magnified. For the same reason, adaptive grids run into the same problem and will miss the correct parameter space region to finer resolve in the first place.

\end{enumerate}

As a consequence of all these shortcomings, it is clear that a method is needed that considers the {\it possibility of systematic differences}. It is also mandatory to consider the finite resolution of our model grids. Solutions to these problems are presented in the following sections.

\subsection{The argument for probabilities}
There are obvious benefits to quantifying the best fit and the uncertainties in terms of probabilities. With probabilities for each specific model, we automatically obtain probability distributions for each parameter of the model itself. We can furthermore consistently compare different grids and see which set of input physics is more probable, given all our current information and the data. 

However, there are much stronger arguments for a probabilistic approach. Marginalization allows us to consistently treat nuisance parameters, while the sum and product rules allow us to clearly formulate the question we are asking. This question is ``Given the observed frequencies, our knowledge about the star and model physics, which model(s) best represent the star in terms of its fundamental parameters and general physical properties as probed by the pulsation modes?" In reality, this general question has to be further refined as we encounter more complicated situations like: ``We have model frequencies that could potentially show negative or positive systematic offsets, or no such offset at all, when compared to our observations. They could be influenced by rotation or actually be rotationally split frequencies themselves. They could be bumped $l=1$ modes or $l=0$ modes. Given all of these possibilities, which model is the most adequate one, and how well is the solution constrained?". From the viewpoint of probability theory the only way to treat such a set of possibilities and get meaningful answers is to use the sum rule and product rule, as we will show in the next section. 

\subsection{Ambiguous mode identification}

As a first improvement to the general approach of asteroseismic model fitting, we can involve the sum rule to consistently consider uncertainties (or even ignorance) in our mode identification. In essence, if there is no unique proposition $f_{i, {\rm o} \mapsto i, {\rm m}}$, Equation\,(\ref{equ:probdens}) changes to
\begin{equation}
P(D | M_j, I) = \prod_{i=1}^{n_{\rm obs}} \left\{ \sum_{k=1}^{n_{\rm match}} P(f_{i, \rm o}, f_{i, {\rm o} \mapsto k, {\rm m}} | M_j, I) \right\}  
\label{equ:probdensnoID}
\end{equation}

with

\begin{equation}
\begin{split}
&P(f_{i, \rm o}, f_{i, {\rm o} \mapsto k, {\rm m}} | M_j, I) = \\
&P(f_{i, {\rm o} \mapsto k, {\rm m}} | M_j, I) P(f_{i, \rm o} | f_{i, {\rm o} \mapsto k, {\rm m}}, M_j, I).
\end{split}
\end{equation}
Here the sum over the index $k$ means that all possible and mutually exclusive assignments $n_{\rm match}$ of one observed mode to a number of calculated frequencies $f_{k, \rm m}$ have to be taken into account as an ``or" proposition\footnote{Hereafter, a possibly ambiguous frequency assignment will always be denoted as $f_{i, {\rm o} \mapsto k, {\rm m}}$.}. Note that due to the product rule of probability, the terms in each sum now include the conditional prior probabilities $P(f_{i, {\rm o} \mapsto k, {\rm m}} | M_j, I)$. These have to be normalized so that $\sum_{k=1}^{n_{\rm match}} P(f_{i, {\rm o} \mapsto k, {\rm m}} | M_j, I) = 1$.  The most conservative assignment is to assign equal probabilities $P(f_{i, {\rm o} \mapsto k, {\rm m}} | M_j, I) = 1/n_{\rm match}$ to each possible scenario. However, if more information is available (e.g., a mode could be identified to be either $l=0$ or $l=2$ with specific probabilities for both cases as found by some peak bagging program), this can easily be encoded at this stage. 

The end result is a product of weighted sums of probabilities, where the weights are given by the respective prior probabilities.\footnote{A common misconception is that these ``priors" are only there to allow us to incorporate prior information. In reality, they are formally required by the product rule and ensure that the result of Equation\,(\ref{equ:probdensnoID}) is always properly normalized.} This product is the correctly normalized likelihood for obtaining the data, given the proposition that any one of the proposed scenarios is correct. Note that if there is an unambiguous assignment $f_{i, {\rm o} \mapsto i, {\rm m}}$ for every observed frequency, each prior probability $P(f_{i, {\rm o} \mapsto i, {\rm m}} | M_j, I) = 1$ and Equation\,(\ref{equ:probdensnoID}) simplifies to Equation\,(\ref{equ:probdens}). Now that we have included our uncertainties concerning the assignment of model frequencies and observed frequencies, we will deal with uncertainties in the validity of the model frequencies themselves in the next section.

\subsection{Treatment of systematic errors}
\label{sec:includesyserr}

As a next step, we now show how to treat the problem of imperfect models. As mentioned before, applying standard techniques that rely on minimizing the quadratic differences between the observations and the models will give incorrect results if systematic differences exist. The alternative of correcting for such imperfections prior to modeling is also undesirable if the correction is not known to be universally applicable. 

To treat any systematic deviation from the model frequencies due to unmodeled physical effects, we simply expand the models $M_j$ by considering an additional systematic error parameter for each tested frequency. The aim is to construct new values $f_{i, \rm \Delta}$ to compare with the observations according to

\begin{equation}
f_{i, \rm \Delta} = f_{i, \rm m} + \gamma \Delta_i
\label{equ:addcorr}
\end{equation}

Here, $\Delta_i$ is the absolute value of the systematic error. $\gamma = 1$ or $\gamma = -1$ and determines whether the model frequency is expected to be systematically higher or lower than the observed frequency. To keep our notation from occupying too much space, we will implicitly assume the value of $\gamma$ to be constant throughout the following derivations, and attribute this to our prior information $I$. $\Delta_i$ is an unknown parameter but as long as its lower and upper boundaries can be roughly estimated, it can be treated fully consistently in the probabilistic framework. 

In the following, we will again work out an example of only one observed and calculated frequency. Therefore, for the derivation the assignment $f_{i, {\rm o} \mapsto i, {\rm m}}$ is unique. We will then provide the extension to multiple frequencies and ambiguous mode identifications. 

Using the new parameters, the equivalent to Equation\,(\ref{equ:singleprob}) is 
\begin{equation}
\begin{split}
&P(f_{i, \rm o}, \Delta_i | f_{i, {\rm o} \mapsto i, {\rm m}}, M^{\Delta}_{j}, I) = \\
&P(\Delta_i | f_{i, {\rm o} \mapsto i, {\rm m}}, M^{\Delta}_j, I) P(f_{i, \rm o} | \Delta_i, f_{i, {\rm o} \mapsto i, {\rm m}}, M^{\Delta}_{j}, I) = \\
&P(\Delta_i | f_{i, {\rm o} \mapsto i, {\rm m}}, M^{\Delta}_j, I) \times \\
&\frac{1}{\sqrt{2\pi}\sigma_{i}}\exp{\left[-\frac{\left(f_{i, \rm o} - f_{i, \rm m} - \gamma \Delta_i \right)^2}{2\sigma_{i}^2}\right]}.
\end{split}
\label{equ:augmented}
\end{equation}

Here the symbol $M^{\Delta}_{j}$ simply denotes the model $M_j$ augmented by the new parameter $\Delta_{i}$.
Self-evidently, the product rule again requires that we introduce a prior probability $P(\Delta_i | f_{i, {\rm o} \mapsto i, {\rm m}}, M^{\Delta}_j, I)$. This can either encode prior information about the expected behavior of the error, or be simply assigned by considerations of symmetry. Again it is required that the integral over the prior $\int P(\Delta_i | f_{i, {\rm o} \mapsto i, {\rm m}}, M^{\Delta}_j, I) \, d\Delta_i = 1$.

It would now be possible to try to find the $\Delta_i$ that maximizes $P(f_{i, \rm o}, \Delta_i | f_{i, {\rm o} \mapsto i, {\rm m}}, M^{\Delta}_{j}, I)$ in Equation\,(\ref{equ:augmented}). However, this is completely irrelevant for our needs. In case of multiple observed frequencies it would also quickly lead to a highly dimensional parameter space that we are not interested in navigating. Instead, we are interested in finding the probabilities of the models $M^{\Delta}_j$. To do this it is necessary to integrate out $\Delta_i$ which we have just introduced. We obtain the marginal likelihood 
\begin{equation}
\begin{split}
&P(f_{i, \rm o} |  f_{i, {\rm o} \mapsto i, {\rm m}}, M^{\Delta}_{j}, I) =  \\ 
&\int_{\Delta_{i, \rm min}}^{\Delta_{i, \rm max}} P(f_{i, \rm o}, \Delta_i | f_{{i, \rm o} \mapsto\ i, \rm m}, M^{\Delta}_{j}, I)  \,d\Delta_i.
\end{split}	
\label{equ:marginalized}			 
\end{equation}
This integral naturally depends on the shape of the prior probability distribution for $\Delta_i$, and can easily be evaluated numerically\footnote{For several simple shapes, such as the {\it beta} prior introduced in the next section, there also exist analytical solutions.}. 
It represents the likelihood of obtaining the value of the observed frequency $f_{i, \rm o}$ given that $M_j$ predicts a frequency $f_{i, \rm m}$ but that there is a possibility of a systematic difference $\Delta_i$, between $\Delta_{i, \rm min}$ and $\Delta_{i, \rm max}$. Furthermore, it is fundamentally constrained and properly weighted by the prior we assigned. This result is now easily extended to multiple modes and ambiguous mode identifications. Equation\,(\ref{equ:probdensnoID}) becomes 
\begin{equation}
P(D | M^{\Delta}_j, I) = \prod_{i=1}^{n_{\rm obs}} \left\{ \sum_{k=1}^{n_{\rm match}} P(f_{i, \rm o}, f_{i, {\rm o} \mapsto k, {\rm m}} | M^{\Delta}_j, I) \right\} 
\label{equ:probdensDeltanoID}
\end{equation}

and

\begin{equation}
\begin{split}
&P(f_{i, \rm o}, f_{i, {\rm o} \mapsto k, {\rm m}} | M^{\Delta}_j, I) = \\
&P(f_{i, {\rm o} \mapsto k, {\rm m}} | M^{\Delta}_j, I) P(f_{i, \rm o} | f_{i, {\rm o} \mapsto k, {\rm m}}, M^{\Delta}_j, I).
\end{split}
\end{equation}
In summary, we have to calculate a product of weighted sums of integrals in the form of Equation\,(\ref{equ:marginalized}), where the summation is performed over every possible assignment $f_{i, {\rm o} \mapsto k, {\rm m}}$.

Note that even when we choose to consider systematic deviations, we usually do not expect them to be significant for all frequencies. For good models some frequencies should already match  well ``right out of the box". In particular, this is true for all frequencies in the idealized case where we have (finally) found a way to correctly model all the effects that previously caused systematic deviations.  

One might think that this is taken care of by setting $\Delta_{i, \min} = 0$. However, unless the prior $P(\Delta_i | f_{i, {\rm o} \mapsto k, {\rm m}}, M^{\Delta}_j, I)$ is a $\delta$ function at $\Delta_i = 0$, it is much more likely that $\Delta_i > 0$. This means that a priori a model will be preferred which shows at least a small deviation from the observations, depending on the observational uncertainties and the steepness of the prior. The limiting case however, the $\delta$ function, corresponds to a whole different model which is simply the standard model without systematic deviations, $M_j$. Thanks to the sum rule, there is an elegant solution for taking this alternative into account. 

For the mutually exclusive logical propositions\footnote{Note that from a logical standpoint even if $\Delta = 0$, $M^{\Delta}_j$ is still a different model than $M_j$ because the prior is not a $\delta$ function. Therefore, they are always mutually exclusive.} $M^{\Delta}_j$ and $M_j$ we can calculate
\begin{equation}
\begin{split}
&P(f_{i, \rm o}, f_{i, {\rm o} \mapsto k, {\rm m}}  | M^{\Delta}_j + M_j, I) = \\
&\frac{P(M^{\Delta}_j, f_{i, \rm o}, f_{i, {\rm o} \mapsto k, {\rm m}} | I) +  P(M_j, f_{i, \rm o}, f_{i, {\rm o} \mapsto k, {\rm m}} | I)}{P(M^{\Delta}_j | I) + P(M_j | I)} = \\
&\frac{P(M^{\Delta}_j | I)}{P(M^{\Delta}_j | I) + P(M_j | I)} P(f_{i, \rm o}, f_{i, {\rm o} \mapsto k, {\rm m}} | M^{\Delta}_j, I) + \\
&\frac{P(M_j | I)}{P(M^{\Delta}_j | I) + P(M_j | I)} P(f_{i, \rm o}, f_{i, {\rm o} \mapsto k, {\rm m}} | M_j, I).
\end{split}
\label{equ:comptwomod}
\end{equation}
Note that here $M^{\Delta}_j + M_j$ means ``$M^{\Delta}_j$ or $M_j$ is true". This is the likelihood of observing the frequency value $f_{i, \rm o}$, given that a systematic deviation either does or does not exist. The principle of indifference as the most conservative approach for the prior probabilities obviously demands $P(M^{\Delta}_j | I) = P(M_j | I) = 0.5$, but if more information is available, it can be encoded here. This result is also easily generalized to the case of multiple frequencies and ambiguous mode identification. 

\subsection{The choice of the prior for $\Delta_i$}
\label{sec:priorchoice}
A very important detail to consider when extending the models with systematic error parameters is their prior probabilities $P(\Delta_i | f_{i, {\rm o} \mapsto k, {\rm m}}, M^{\Delta}_j, I)$. There is a basic choice between two possibilities. The first is to use uninformative (or ignorance) priors, or alternatively, maximum entropy priors. Uninformative priors can be derived from arguments of invariance to specific transformations, while maximum entropy priors should satisfy the maximum entropy criterion for a given set of constraints. The other possibility is to use priors derived from heuristic or physical arguments. 

The specific form of the prior probabilities of $\Delta_i$ are part of the model that is evaluated, as indicated by the notation.\footnote{The prior is described by $P(\Delta_i | f_{i, {\rm o} \mapsto k, {\rm m}}, M^{\Delta}_j, I)$ rather than, e.g., $P(\Delta_i | I)$} They are not necessarily ``prior" as in a sense of ``before obtaining observations", but conditional probabilities required for the correct normalization, as demanded by the product rule of probabilities. They encode specific ways in which we expect $\Delta_i$ to behave, given our grid of frequencies and our information (which of course can be influenced by previous observations). For instance, if we expect our best model to minimize the systematic deviations, the prior should assign larger probability densities to smaller $\Delta_i$, so that models with smaller deviations will be more probable. On the other hand, if we expect our best model to show more erratic deviations, a flat uninformative prior is a better choice. After a complete evaluation of the probabilities and likelihoods, the Bayesian evidence will indicate whether the state of information encoded by the priors is supported by the data or not. 

As a first important example of an uninformative prior, consider a uniform prior  
\begin{equation}
P(\Delta_i | f_{i, {\rm o} \mapsto k, {\rm m}}, M^{\Delta}_j, I) = \frac{1}{\Delta_{i,\rm max} - \Delta_{i,\rm min}} = {\rm const}.
\label{equ:uniform}
\end{equation}
This means that all values of $\Delta_i$ are equally likely. With such a prior, every model that predicts frequencies at any value between 
$f_{i, \rm o} + \Delta_{i, \rm min}$ and $f_{i, \rm o} + \Delta_{i, \rm max}$ has the same maximum likelihood 
(i.e., the same maximum value for Equation\,(\ref{equ:augmented})). 

On the other hand, a Jeffreys prior 
\begin{equation}
P(\Delta_i | f_{i, {\rm o} \mapsto k, {\rm m}}, M^{\Delta}_j, I) = \frac{1}{\Delta_{i} \ln\left(\Delta_{i,\rm max}/\Delta_{i,\rm min}\right)},
\label{equ:jeffreys}
\end{equation}
assigns equal probability per decade and, in terms of the probability density, favors smaller values of $\Delta_{i}$. This prior is obviously not defined for $\Delta_{i} = 0$, i.e., it requires $\Delta_{i, \rm min} > 0$. This is problematic for, e.g., surface effects that approach zero at low orders. However, when $\Delta_{i, \rm min} = 0$, one can use a modified version of this prior given by
\begin{equation}
P(\Delta_i | f_{i, {\rm o} \mapsto k, {\rm m}}, M^{\Delta}_j, I) = \frac{1}{\left(\Delta_{i} + c\right) \ln\left[\left(\Delta_{i,\rm max} + c\right) / c\right]},
\label{equ:modjeffreys}
\end{equation}
where $c$ is a small constant. For values smaller than $c$, this prior acts more or less like a uniform prior, while for higher values it behaves like the usual Jeffreys prior. This prior is nowadays often used in ``peak-bagging" algorithms \citep[e.g.,][]{gruberbauer09, benomar2009, handberg11}. However, there is no objective criterion for how to set $c$, and various tests we conducted with our grid fitting code have shown that the choice of $c$ can have a large impact on the evidence values. 

Consequently, we argue that any priors used for a systematic error parameter $\Delta_{i} = [0, \Delta_{i, \rm max}]$ should be functions that are clearly defined by the parameter limits, without additional parameters that have large effects on the evidence. The uniform prior\footnote{In fact, the uniform prior is consistent with a beta distribution with $\alpha = \beta = 1$.} is such a prior, as are priors derived from the beta distribution (given in units of our problem)

\begin{equation}
P(\Delta_i | f_{i, {\rm o} \mapsto k, {\rm m}}, M^{\Delta}_j, I) \propto \left(\frac{\Delta_i}{\Delta_{i, \rm max}}\right)^{\alpha - 1} \left(1 - \frac{\Delta_i}{\Delta_{i, \rm max}}\right)^{\beta - 1}.
\label{equ:betadist}
\end{equation}

With $\alpha = 1$ and $\beta = 2$ this simply leads to a linearly decreasing probability density 

\begin{equation}
P(\Delta_i |f_{i, {\rm o} \mapsto k, {\rm m}}, M^{\Delta}_j, I) = \left(\frac{\sqrt{2}}{\Delta_{i, \rm max}}\right)^2 \left(\Delta_{i, \rm max} - \Delta_i\right).
\label{equ:betaprior}
\end{equation}

It is the only prior that allows for a linearly decreasing probability density, is properly normalized, and reaches zero at $\Delta_i = \Delta_{i, \rm max}$. It also leads to an analytical solution for the integral in Equation\,(\ref{equ:marginalized}). Thus, it satisfies all our requirements for a prior with which to minimize systematic errors. 

We have compared the results obtained from Equation\,(\ref{equ:betaprior}) (hereafter: {\it beta} prior) with several other plausible choices, such as an exponential distribution with expectation value $\Delta_{i, \rm max}/2$ and a modified Jeffreys prior with $c = \sigma_{i, \rm m}$, and we find them to yield comparable results and evidence values. Due to the clarity of its definition and lack of additional parameters, we therefore argue that the {\it beta} prior is an appropriate choice for a non-flat prior. We will show how to use it in Section\,\ref{sec:surface} in a worked example.

Lastly, priors based on heuristic or physical arguments obviously vary strongly with the specific problem to which the fitting method is applied. As an example, when modeling surface effects on {\it p}-mode frequencies, the prior could be Gaussian, following the heuristic frequency correction proposed by \cite{kjeldsen2008}. It would be a function of frequency, expecting greater deviations toward higher-order modes. The width of the Gaussian, however, would be again a rather arbitrary choice, leading to potentially different evidence values. Such priors clearly need to have a strong basis either in theory or prior observations.

\subsection{Bumped modes and finite grid resolution}
Equation\,(\ref{equ:probdensDeltanoID}) represents the final likelihood for obtaining the observed frequencies given our (extended) model $M^{\Delta}_j$.  
This model still represents only a single point in a discrete grid.\footnote{Note that this is also the case for approaches using an adaptive grid, since each iteration of an adaptive scheme is based on a discrete grid.} However, the probability is small that a single model in the grid corresponds to the ``true best model" our code can produce. The problem becomes worse as the grid resolution is lowered, or as mode frequencies are changing quickly {\it or unpredictably} from one model to the next (e.g., avoided crossings, magnetic shifts). The probabilities (or $\chi^2$-values) we obtain will not be a fair assessment of the model physics, even at higher grid resolutions. Even worse, the overall evidence for the grid will be finely tuned to the positions of all models in the grid. This makes it difficult to compare different grids with different physics. We will now show how to improve on this.

In a sequence of models along a single evolutionary track, except for the first and last models, each model $M_j$ has two neighboring models $M_{j-1}$ and $M_{j+1}$. In most cases these adjacent models will contain the same modes, and their changing values can be traced from $M_{j-1}$ to $M_j$ and $M_{j+1}$. Now we declare the difference between observed and calculated frequency as a new free parameter
\begin{equation}
\delta f_{i} = f_{i, \rm o} - f_{i, \rm m}.
\end{equation}

This value is fixed if only a single grid point is considered. However, we can split the evolutionary tracks
into segments in between grid points, and define

\begin{equation}
\delta f_{i, j-} = f_{i, \rm o} - \frac{f_{i, M_{j-1}} + f_{i, M_{j}}}{2}
\end{equation}

and equivalently

\begin{equation}
\delta f_{i, j+} = f_{i, \rm o} - \frac{f_{i, M_{j}} + f_{i, M_{j+1}}}{2}.
\end{equation}

Adding $\delta f_i$ as a new parameter to the equations derived in the earlier sections, we change our focus to evaluate probabilities of model track segments $T^{\Delta}_{j}$ centered around the models $M^{\Delta}_{j}$. To do this, we again use marginalization to integrate out both $\Delta_i$ and $\delta f_i$ to obtain the marginal likelihood. We obtain
\begin{equation}
\begin{split}
&P(f_{i, \rm o} |  f_{i, {\rm o} \mapsto i, {\rm m}}, T^{\Delta}_{j}, I) =  \\ 
&\int_{\Delta_{i, \rm min}}^{\Delta_{i, \rm max}} \int_{\delta f_{i, j-}}^{\delta f_{i, j+}} P(f_{i, \rm o}, \Delta_i, \delta f_i | f_{i, {\rm o} \mapsto i, {\rm m}}, T^{\Delta}_{j}, I)  \,d\Delta_i \,d\delta f_i.
\end{split}	
\label{equ:erf}			 
\end{equation}

If the priors for $\Delta_i$ do not vary greatly from one model to the next, $\Delta_i$ and $\delta f_i$ can be considered to be independent parameters. It is therefore possible to use the product rule to separate the conditional probabilities

\begin{equation}
\begin{split}
P(f_{i, \rm o}, &\Delta_i, \delta f_i | f_{i, {\rm o} \mapsto i, {\rm m}}, T^{\Delta}_{j}, I) = \\
&P(\Delta_i |  f_{i, {\rm o} \mapsto i, {\rm m}}, T^{\Delta}_{j}, I) \times \\ 
&P(\delta f_i | f_{i, {\rm o} \mapsto i, {\rm m}}, T^{\Delta}_{j}, I) \times \\
&P(f_{i, \rm o} | \Delta_i, \delta f_i, f_{i, {\rm o} \mapsto i, {\rm m}}, T^{\Delta}_{j}, I). 
\end{split}	
\label{equ:erf2}			 
\end{equation}

\begin{figure}[!ht]
\centering
\includegraphics[width=0.9\columnwidth]{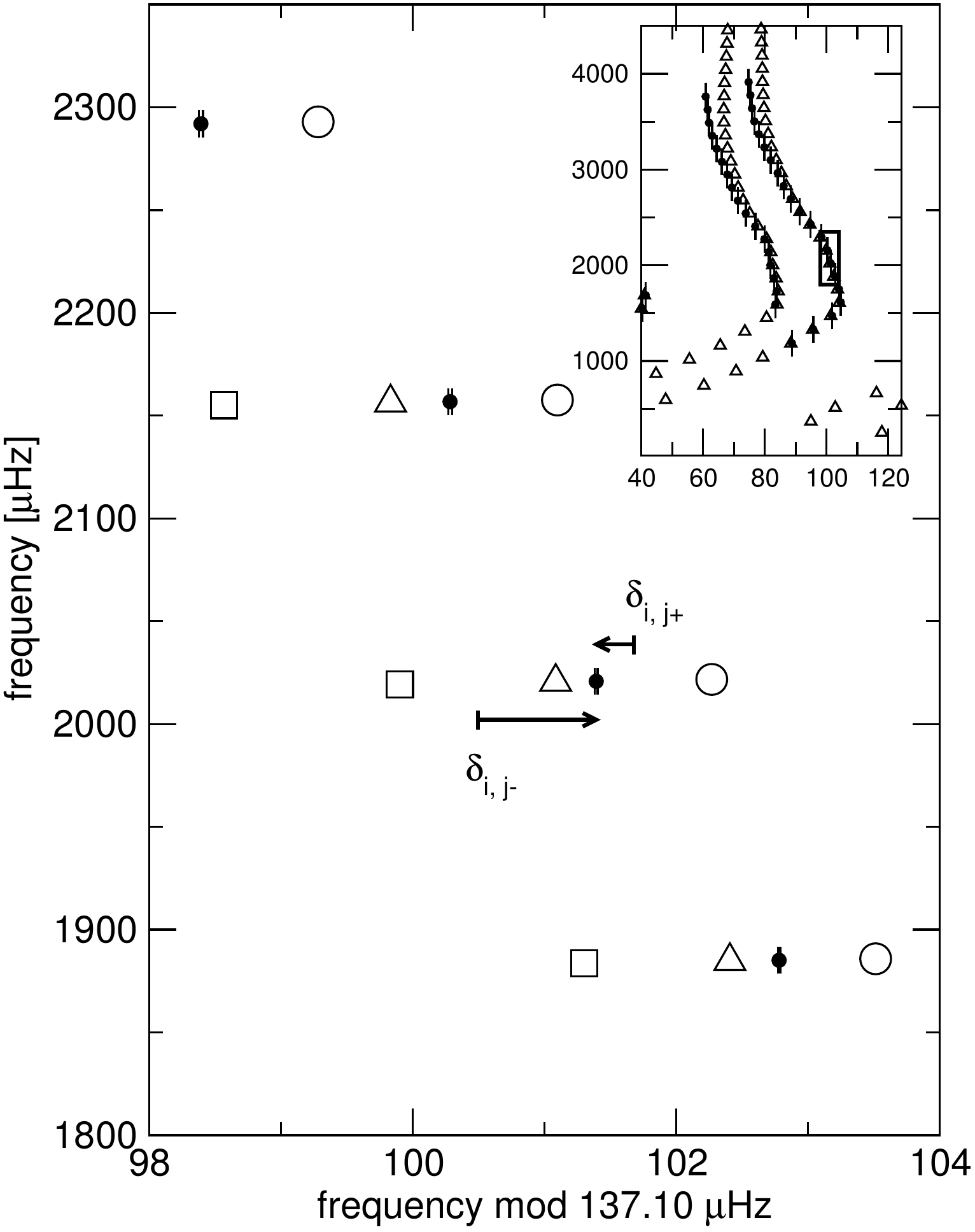}
\caption{Example for the definition of $\delta_{i, j-}$ and $\delta_{i, j+}$ (see the text). Four radial orders of $l=1$ modes from three adjacent models in a high-resolution grid of solar models are shown. The triangles represent the central model. The frequencies for the adjacent models along the evolutionary track sequence are depicted as squares and white circles. Black circles (and error bars) indicate observed frequencies published in \cite{broomhall2009}. $\delta_{i,j-}$ and $\delta_{i,j+}$ for a single mode are represented using arrows.  The insert shows an unzoomed version of the $l=1$ and $l=3$ ridge.}  
\label{fig:erfplot}
\end{figure}

Furthermore, since we evaluate the complete evolutionary track segment we can assume a uniform prior probability $P(\delta f_i |  f_{i, {\rm o} \mapsto i, {\rm m}}, T^{\Delta}_{j}, I) = 1/\left(\delta f_{i, j+} - \delta f_{i, j-}\right)$. With these definitions, the integral over $\delta f_i$ can easily be calculated analytically. The equivalent to Equation\,(\ref{equ:augmented}) becomes

\begin{equation}
\begin{split}
&P(f_{i, \rm o}, \Delta_i | f_{i, {\rm o} \mapsto i, {\rm m}}, T^{\Delta}_{j}, I) = \\
&P(\Delta_i | f_{i, {\rm o} \mapsto i, {\rm m}}, T^{\Delta}_{j}, I) \times \\
&\frac{1}{2\left(\delta f_{i, j+} - \delta f_{i, j-}\right)}\times \\
&\left[{\rm erf} \left(\frac{\delta f_{i, j+} - \gamma \Delta_i}{\sqrt{2}\sigma_i}\right) - {\rm erf}\left(\frac{\delta f_{i, j-} - \gamma \Delta_i}{\sqrt{2}\sigma_i}\right)\right].
\end{split}
\label{equ:augmentederf}
\end{equation}
where ${\rm erf}$ is the error function.
The remaining integral over $\Delta_i$ again has to be carried out numerically.
Figure\,\ref{fig:erfplot} shows an example for the definitions introduced above, given three models in a solar evolutionary track. 

We have now used the free parameter $\delta f_i$ to ``trace" each mode through segments of the evolutionary track, and compare it to the observed frequencies, retaining the possibility of systematic differences. Note that our only assumption here is that the mode frequencies change smoothly between the frequencies given by the constraining models. In principle, this approach can be carried out in multiple dimensions (e.g., not only along the evolutionary track in stellar age but also between tracks in mass). As before, an extension to multiple frequencies and ambiguous mode identifications is straightforward. 

We stress that this approach only locates the region of highest probability given the current grid, and given unspecified behavior of frequencies in between grid points. It is thus best used for frequencies whose behavior is difficult to capture, e.g., due to mode bumping, or for a first general assessment of a very coarse grid. Given a dense enough grid, regular frequencies that are expected to change approximately linearly from one grid point to the next need to be treated using interpolation, since the integration over the model gaps for individual modes, independently of all other modes, would allow for highly unphysical models. 

Therefore, in order to obtain a final best model and uncertainties for the model parameters, the regions of substantial probability should be further refined after the track probabilities have been calculated. Eventually, the grid is resolved enough so that well-defined distributions arise. In dense enough grids, this can easily be accomplished by interpolation of the frequencies in between grid points without violating the condition of hydrostatic equilibrium. This can also be done during run-time with arbitrary precision using probabilities, by interpolating between grid points and using the sum rule to calculate a probability representative of the original grid resolution using the interpolated models. Naturally, modes that change erratically, should be excluded from such an interpolation routine and treated as shown above instead.

\subsection{Model probabilities}
So far we have only shown how to calculate the {\it likelihood} for standard pulsation models, models that contain systematic differences, and also evolutionary track segments. In order to obtain the {\it probabilities} for individual models (or track segments), we want to use Bayes' theorem, assign model priors, and calculate the total evidence for each model grid. The simplest method to assign model priors in the absence of any other prior information, is to use the principle of equipartition and assign a uniform prior
\begin{equation}
P(M_j | I) = 1.0/N_{\rm M},
\end{equation}
where $N_{\rm M}$ is the number of models (or, equivalently, $N_{\rm T}$ would be the number of track segments) that are analyzed.

Although each model or track only predicts a number of frequencies, it implicitly represents values or ranges for fundamental parameters like $T_{\rm eff}$ or $L$, which can be compared to (and constrained by) different and non-seismic observations. For instance, assuming our prior photometric and spectroscopic observations of a pulsating star indicate $T_{\rm eff} = T_{\rm spec} \pm \sigma_{\rm spec}$ then the prior probability density for the model temperature is
\begin{equation}
P(T_{{\rm eff},j} | I) = k\exp{\left[-\frac{\left(T_{\rm spec} - T_{{\rm eff},j}\right)^2}{2\sigma_{\rm spec}^2}\right]}. 
\end{equation}
This example assumes that the uncertainty in $T_{\rm eff}$ follows a Gaussian distribution. $k$ is a normalization constant depending on the absolute lower and upper plausible limits of $T_{\rm eff}$. 

All the different implicit parameters for which prior observations or other fundamental constraints are available, and hence prior knowledge exists, then can be used for prior probabilities which combine into an overall prior for model $M_j$. As an example
\begin{equation}
P(M_j | I) = P(T_{{\rm eff},j} | I) P(L_j | I) P(\left[\rm Fe/H\right]_j | I) ... 
\end{equation}
if we assume separable priors for simplicity. If probabilities of track segments are calculated, such a prior could be approximated by a product of separable integrals, which are easily evaluated analytically. Again, in order to obtain a proper prior and therefore proper values for the evidence, the integral of the prior probability over all possible models/tracks in a grid should be 1. 

By calculating, e.g., 
\begin{equation}
\begin{split}
P(M^{\Delta}_j | D, I) = \frac{P(M^{\Delta}_j | I) P(D | M^{\Delta}_j, I)}{\sum_{k=1}^{N_{\rm M}} P(M^{\Delta}_k| I) P(D | M^{\Delta}_k, I)}
\end{split}
\label{equ:compmod}
\end{equation}
or, e.g., in the case of rapidly changing modes with or without systematic errors, 
\begin{equation}
\begin{split}
P(T^{\Delta}_{j}  + T_j | D, I) = \frac{P(T^{\Delta}_{j} + T_j | I) P(D | T^{\Delta}_{j}  + T_j, I)}{\sum_{k=1}^{N_{\rm T}} P(T^{\Delta}_{k}  + T_j | I) P(D | T^{\Delta}_{k}  + T_j, I)}
\end{split}
\label{equ:comptrack}
\end{equation}
we obtain the probability of $M^{\Delta}_j$ or, respectively, $T^{\Delta}_{j} + T_j$ given our prior knowledge (or lack thereof), our grid, and the set of observed frequencies. Note that the denominators of these equations represent the evidence or likelihood for the grid as a whole. We can therefore use these as {\it likelihoods} when we want to compare different grids with different input physics.


\section{Application to surface effects}
\label{sec:surface}

As mentioned in the introduction, shortcomings in modeling the outer stellar layers produce systematic deviations in comparison to the observations. These deviations seem to be such that model frequencies tend to be higher than the observed frequencies, and therefore $\gamma = -1$ (see Equation\,(\ref{equ:addcorr})). \cite{kjeldsen2008} have proposed to calibrate a power-law description of the deviations by measuring the surface effects in the Sun, and then fitting this relation to frequencies of other stars. Their correction expressed in terms of our definitions has the general form of
\begin{equation}
\gamma \Delta_i \approx a \left(\frac{f_{i, \rm m}}{f_{\rm ref}}\right)^{b},
\label{equ:kjeldsen}
\end{equation}
where $f_{\rm ref}$ is some reference frequency, typically the frequency of maximum power $\nu_{\rm max}$, and $a$ and $b$ are parameters to be fitted. From their fits, \citeauthor{kjeldsen2008} determined $b \approx 4.90$ in the Sun, which has subsequently been used for other stars by a number of authors. A comprehensive implementation of this formalism into a $\chi^2$-fitting algorithm was presented in a recent study by \cite{brandao2011}. However, even in this more advanced approach, there is still a choice of $a$ and $b$ required. Moreover, complications for modes of different spherical degree and also bumped modes arise because they do not necessarily conform to this relation. The authors propose to alleviate these problems by introducing additional model-dependent parameters that approximately correct for some of these deviations. While this approach is a great improvement over applying a fixed surface-effect correction (or no correction at all), our approach is much more powerful. It allows for a much greater flexibility and leads to clearly defined probabilistic results.

\begin{figure*}[t!]
   \centering
	   \includegraphics[width=0.45\textwidth]{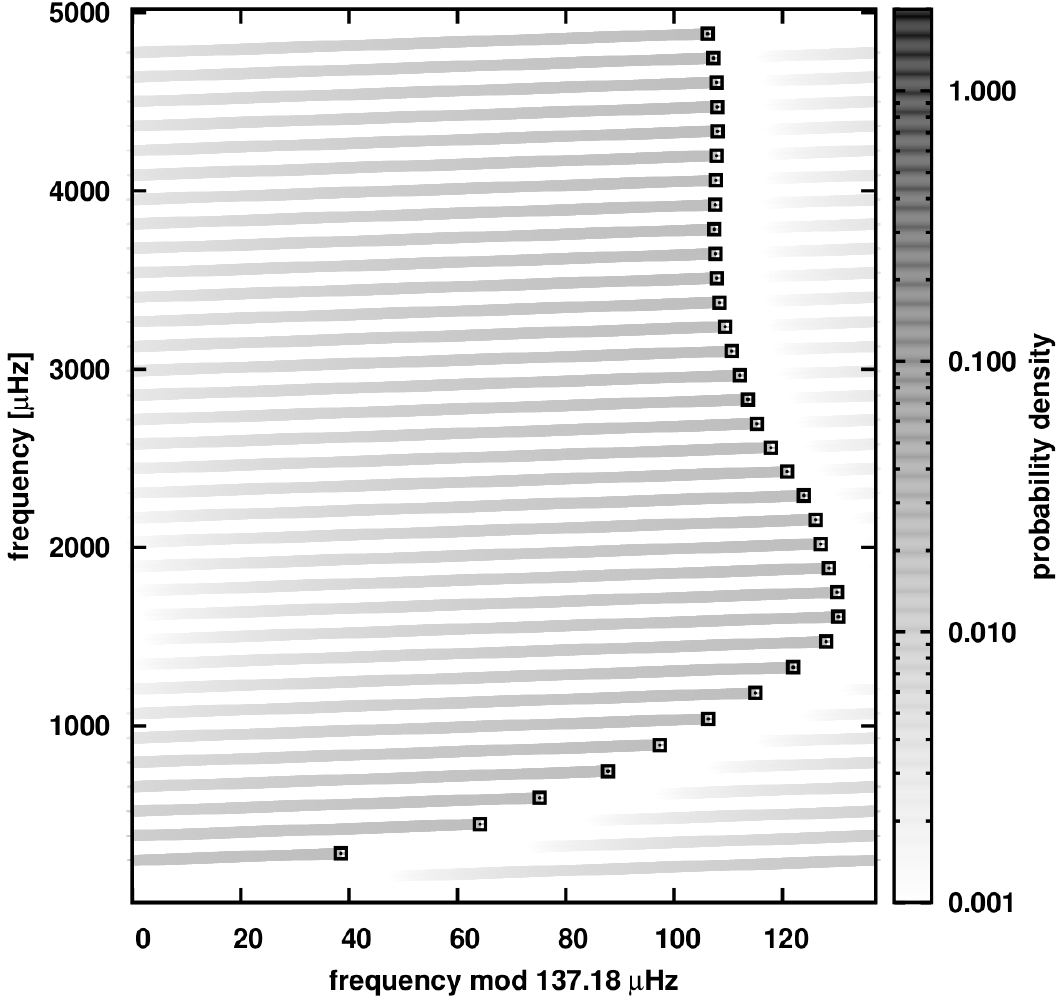}
  	   \includegraphics[width=0.45\textwidth]{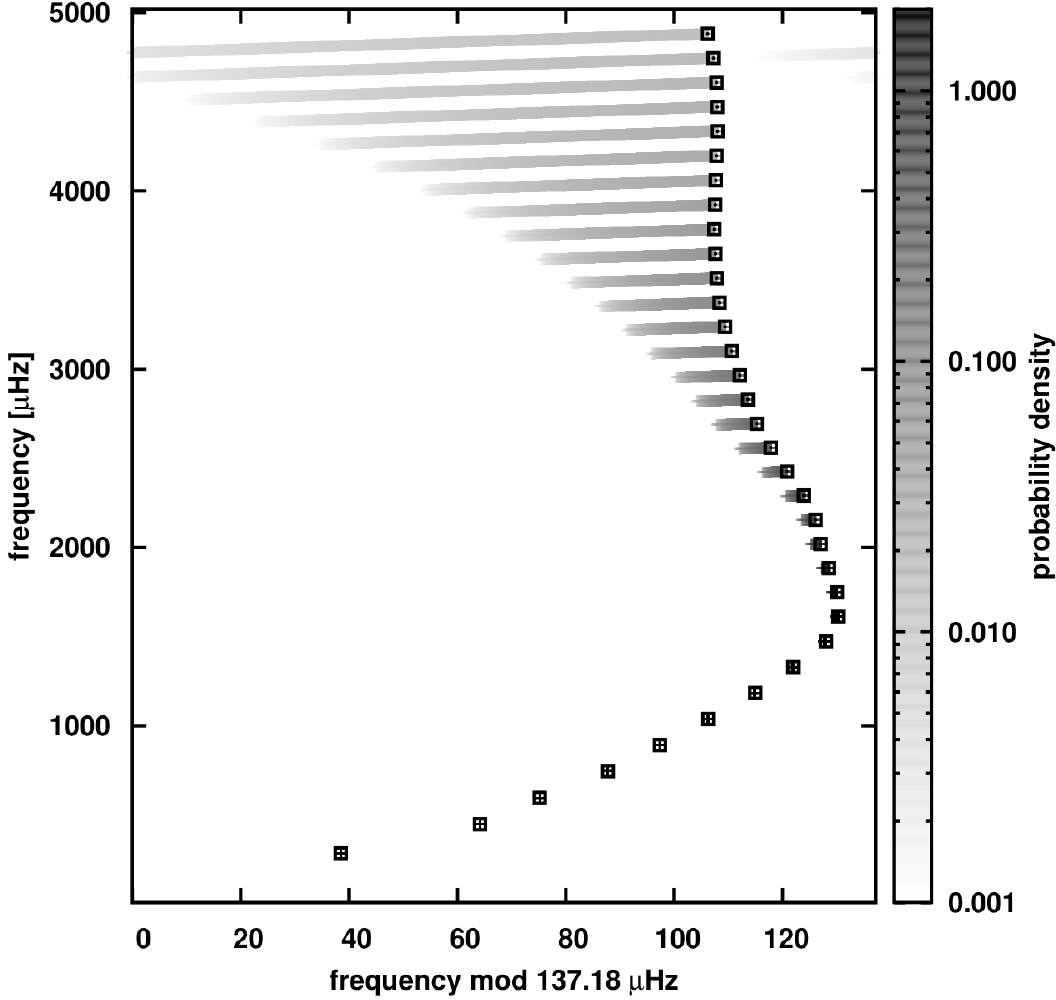}
	   \caption{Behavior of the {\it beta} prior for systematic offsets in an echelle diagram. The squares represent model frequencies, while the shaded ``trails" indicate the prior probability density for varying $\Delta_i$. The left panel uses equal $\Delta_{\rm max}$, whereas the right panel uses a power-law $\Delta_{\rm max}$ with exponent $b = 4.9$ (see Section\,\ref{sec:surfprior}). Note that the uniform prior is not shown, since it simply assigns a constant probability density.}
   \label{fig:probsurf}
\end{figure*}

\subsection{Priors for surface effects}
\label{sec:surfprior}

As we want to treat systematic errors of more or less unknown magnitude, the most general approach is to use the flat uniform prior (Equation\,(\ref{equ:uniform})). Imposing only minor additional constraints, as we argued in Section~\ref{sec:priorchoice}, the {\it beta} prior (Equation\,(\ref{equ:betaprior})) can also be used to give more weight to models which minimize these unknown errors. We can use both priors and compare the Bayesian evidence to tell us which interpretation of the surface effect is better supported by the data, given our model and everything we know. Moreover, irrespective of which prior is chosen, we also always allow for the possibility of no surface effects at all, as discussed in Section\,\ref{sec:includesyserr}.

This now gives us enough flexibility to consider a possibly frequency-dependent behavior of the surface effects. However, instead of ``predicting" the behavior of $\Delta_i$ as is done by modeling the surface deviations through a power law, we will prescribe the behavior of its {\it upper limit} $\Delta_{i, \rm max}$. In contrast, the lower limit should always remain 0, since our ultimate goal is to find models that correctly describe the surface layers  (and therefore approach $\Delta_i \rightarrow 0$). 

The choice of the largest allowed $\Delta_{\rm max}$ is not unique, but it should be sensible and used consistently throughout the analysis. A reasonable strategy is to use $\sup(\Delta_{\rm max}) = \Delta\nu$, the large frequency separation of each specific model, as a sensible upper limit. If the systematic differences between observations and models are larger than the average distance between modes of adjacent radial order, we no longer recognize this as a valid frequency assignment\footnote{This condition may be relaxed at the highest radial orders.}. With this upper limit defined, we now want to model different types of surface effects. If we have no preference for any frequency-dependent trend (i.e., all we know is that observed frequencies are lower than model frequencies) we require that all frequencies have equal $\Delta_{\rm max}$. 

On the other hand we can also use a more specific model, such as Equation\,(\ref{equ:kjeldsen}), but retain the same flexibility. The surface effect as shown in Equation\,(\ref{equ:kjeldsen}) depends on two parameters. The power-law exponent $b$ determines how quickly the surface effect increases as we move to higher frequencies, whereas $a$ is simply a scaling factor. We are not interested in the scaling parameter, since the scaling (i.e., the magnitude of the offset) is governed by our condition that for each model $\sup(\Delta_{\rm max}) = \Delta\nu$. It is taken care of by the fact that we are marginalizing over $\Delta_i$ anyway. Since the largest surface effects are expected at the highest frequency $f_{\rm max, m}$ in the model, it follows that for a specific $b$ 
\begin{equation}
\Delta_{{\rm max}, i} = \Delta\nu \left(\frac{f_{i, \rm m}}{f_{\rm max, m}}\right)^b.
\label{equ:dNuscaling}
\end{equation} 
Figure\,\ref{fig:probsurf} shows how these definitions affect the prior probability density $P(\Delta_i | f_{i, {\rm o} \mapsto i, {\rm m}}, M^{\Delta}_j, I)$ as we increase the value of $\Delta_i$ for both the constant and the power-law approach. With all the $\Delta_{{\rm max}, i}$ set, we can then use the priors as discussed above for all our calculations. Note that we can also easily evaluate new composite propositions at this stage and compute the probability for a hypothesis that allows for, e.g., a range $b = \{4.4, 4.65, 4.9, ...\}$. This is done in the same way as was explained earlier (see Equation\,(\ref{equ:comptwomod})).

\subsection{Detailed analysis of the Sun}

As an example for how to implement the surface-effect treatment, we will consider the solar $l=0$, 1, 2, and 3 {\it p} modes obtained by using BiSON data \citep{broomhall2009}. 
For our models, we used a large and dense solar grid obtained with YREC \citep{demarque2008}. The model grid spans: masses from 0.95 $M_{\odot}$ to 1.05 $M_{\odot}$ in steps of 0.005 $M_{\odot}$, initial hydrogen mass fractions from 0.68 to 0.74 in steps of 0.01, initial metal mass fractions from 0.016 to 0.022 in steps of 0.001, and mixing length parameters from 1.8 to 2.4 in steps of 0.1. These parameters are also summarized in Table\,\ref{tab:sungridpar}. 

Our model tracks begin as completely convective Lane--Emden spheres \citep{lane1869, chandra1957} and are evolved from the Hayashi track \citep{hayashi1961} through the zero-age main sequence (ZAMS) to 6 Gyr with each track consisting of approximately 2500 models. Only models between 4.0 and 6.0 Gyr are included in the model grid. Constitutive physics include the OPAL98 \citep{iglesias1996} and \cite{alexander1994} opacity tables using the GS98 mixture \citep{grevesse1998}, and the Lawrence Livermore 2005 equation of state tables \citep{rogers1986, rogers1996}. Convective energy transport was modeled using the B\"ohm-Vitense mixing-length theory \citep{boehm1958}. The atmosphere model follows the ($T$--$\tau$) relation by \cite{krishna1966}. Nuclear reaction cross-sections are from \citep{bahcall2001}. The effects of helium and heavy element diffusion \citep{bahcall1995} were included. Note that our atmosphere models and diffusion effects have been shown to require a larger value of mixing length parameter ($\alpha_{\rm ml} \approx 2.0 -- 2.2$) than standard Eddington atmospheres ($\alpha_{\rm ml} \approx 1.7 -- 1.8$) \citep{guenther1993}. 

The pulsation spectra were computed using the stellar pulsation code of \cite{guenther1994}, which solves the linearized, non-radial, non-adiabatic pulsation equations using the Henyey relaxation method. The non-adiabatic solutions include radiative energy gains and losses but do not include the effects of convection. We estimate the random $1\sigma$ uncertainties of our model frequencies to be of the order of $0.1\,\mu\rm Hz$.

\begin{table}[h!]
   \centering
   \caption{Parameter Ranges for the Solar Grid.}
   \begin{tabular}{c l l } 
   \hline
   \hline
      Parameter & Range & Step Size \\
   \hline 
    Mass & 0.95--1.05 & 0.005 \\
    $X_0$ & 0.68--0.74 & 0.01 \\
    $Z_0$ & 0.016--0.022 & 0.001 \\
    $\alpha_{\rm ml}$ & 1.8--2.4 & 0.1 \\     
     \hline     
   \end{tabular}
   \tablenotetext{0}{\textbf{Notes.} Masses are given in units of solar masses; $\alpha_{\rm ml}$ is the mixing length parameter.}
   \label{tab:sungridpar}
\end{table}

We analyzed our grid using adiabatic and non-adiabatic frequencies, and employed three different surface-effect models: 
\begin{itemize}
\item \textbf{M1}: frequency-independent surface effects with $\Delta_{i, \rm max} = \Delta\nu$ 
\item \textbf{M2}: frequency-dependent, ``canonical" surface effects with $\Delta_{i, \rm max}$ following Equation\,(\ref{equ:dNuscaling}) with $b = 4.90$
\item \textbf{M3}: same as \textbf{M2}, but with $b$ as a free parameter marginalized from $b=3.0$ to $b=6.0$
\end{itemize}

For each frequency evaluated throughout our model grid, irrespective of the surface-effect model, we also considered the possibility of no surface effect, i.e., we consistently calculated $P(M^{\Delta}_j + M_j | D, I)$. To take into account the discrete nature of the grid, we interpolated along the evolutionary tracks during run-time by a factor of 20, thereby increasing the effective ``frequency resolution" of the grid to below the random uncertainties of the model frequencies.
All models were evaluated with 
\begin{itemize}
\item (\textbf{a}) a uniform prior for all track segments
\item (\textbf{b}) a prior using normal distributions for the constraints $M = 1.0000 \pm 0.0002\, M_{\odot}$, $\log T_{\rm eff} = 3.7617 \pm 0.01$, and $\log\left(L/L_{\odot}\right) = 0.00 \pm 0.01$, where YREC uses the following adopted values for $M_{\odot} = 1.9891 \pm 0.0004 \cdot 10^{33}\rm g$ \citep{cohen1986} and $L_{\odot} = 3.8515 \pm 0.009 \cdot 10^{33} \rm erg\,s^{-1}$ (the average of the ERB-Nimbus and SMM/ARCRIM measurements; \cite{hickey1983})
\item (\textbf{c}) same as (\textbf{b}) but with an additional Gaussian constraint on the age of $\rm 4.603 \pm 0.0075 \rm\,Gyr$, derived from the estimated age of the solar system found by \cite{bouvier2010} and an average pre-main sequence phase of our models of $35\pm 5\rm \,Myr$.
\end{itemize}

For the $\Delta_i$ we consistently used {\it beta} priors, as discussed in the previous section. Our calculations yield the most probable models and uncertainties for all these approaches, and they also give the Bayesian evidence for each approach. The results are summarized in Table\,\ref{tab:sunresults}. We also computed the probabilities using uniform priors, but found similar results with lower evidence (several orders of magnitude) values than for the corresponding {\it beta} prior analysis. 

\begin{table}[ht!]
   \centering
   \caption{Evidence for the Solar Grid Using the BiSON Data Set}
    \begin{tabular}{l l l l} 
   \hline
   \hline
     Surface & HRD  & $\log_{10}(evidence)$ & $\log_{10}(evidence)$ \\
     Model & Prior &  (adiabatic) & (non-adiabatic) \\
   \hline 
	\textbf{M1} & \textbf{a} &$ -233.4$ & $-229.9$ \\
	\textbf{M2} & \textbf{a} & $-189.8$ & \textbf{-186.7} \\
	\textbf{M3} & \textbf{a} & $-189.8$ & $-187.2$ \\
	\hline 
	\textbf{M1} & \textbf{b} &$ -235.5$ & $-231.6$ \\
	\textbf{M2} & \textbf{b} & $-190.8$ & \textbf{-187.1} \\
	\textbf{M3} & \textbf{b} & $-191.4$ & $-187.9$ \\
	\hline
	\textbf{M1} & \textbf{c} &$ -236.7$ & $-235.1$ \\
	\textbf{M2} & \textbf{c} & $-193.5$ & \textbf{-190.7} \\
	\textbf{M3} & \textbf{c} & $-192.6$ & $-189.4$ \\
    \hline     
   \end{tabular}
   \tablenotetext{0}{\textbf{Notes.}  See the text for the definition of models and prior a, b, and c. Results for models \textbf{M2a}, \textbf{M2b}, and \textbf{M2c}, which are analyzed in more detail, are indicated in bold face. Note that small numbers are expected.}
   \label{tab:sunresults}
\end{table}

\begin{table*}[tbp]
   \centering
   \caption{Most Probable Models for the Complete BiSON Data Set Model Fitting.}
    \begin{tabular}{l c c c c c c c} 
   \hline
   \hline
     Model &  Mass & Age & $X_0$ & $Z_0$ & $Z_{\rm s}$ & $\alpha_{\rm ml}$  & Probability \\
     \hline
     \textbf{M2a} & $ 1.015 $ & $4.885 \pm 0.006$ & 0.73 & 0.017 & 0.0153 & 2.2 & 0.54 \\
             & $ 1.005 $ & $4.713 \pm 0.006$ & 0.72 & 0.017 & 0.0153 & 2.2 & 0.21 \\
    \hline
     \textbf{M2a} & $ 1.000 $ & $5.017 \pm 0.006$ & 0.72 & 0.018 & 0.0161 & 2.1 & 0.68 \\
     	     &	$ 1.000 $ & $4.983 \pm 0.006$ & 0.71 & 0.019 & 0.0160 & 2.2 & 0.17 \\
     \hline        
     \textbf{M2a} & $1.000$ & $4.591 \pm 0.005$ & 0.72 & 0.016 & 0.0144 & 2.2 & 0.95 \\
             & $1.000$ & $4.562 \pm 0.005$ & 0.71 & 0.017 & 0.0153 & 2.3 & 0.05 \\ 
    \hline     
   \end{tabular}
   \tablenotetext{0}{\textbf{Notes.} Age is given in billion years and is computed from the pre-main-sequence birthline. The age from the ZAMS is $35 \pm 5$ million years less. $X_0$ and $Z_0$ are initial hydrogen and metal mass fractions, $Z_{\rm s}$ is the metal mass fraction in the envelope. Probabilities are given with respect to the specific surface-effect model and prior combination.}
   \label{tab:sundist}
\end{table*}

The non-adiabatic frequencies consistently produce larger evidence values than for the respective adiabatic case. This is no surprise, as the non-adiabatic frequencies are in general better at reproducing the higher frequencies. 
Overall, model \textbf{M2a} shows the largest evidence, followed by \textbf{M2b} and \textbf{M3a}. Note that \textbf{M1a}, \textbf{M1b}, and \textbf{M1c}, which use frequency-independent priors for the surface effects, and therefore are extremely flexible, fail compared to the other models. Also, \textbf{M3a} and \textbf{M3b} cannot beat their \textbf{M2} counterparts. These are examples of how {\it marginalization and the consistent normalization of probabilities work together to penalize more flexible models if they cannot generate considerably better results}. Model M3c has a greater evidence than M2c, but the most probable stellar models are the same in both cases, suggesting that these models fit well, but do not necessarily adhere in detail to the standard surface correction.

At first glance it might be unsettling that \textbf{M2a} has a slightly greater evidence than \textbf{M2b} (and significantly greater evidence than \textbf{M2c}). This indicates that there are models in our grid which reproduce the pulsation spectrum very well but do not match the solar fundamental parameters. A correctly calibrated grid would produce higher evidences with a prior restricted to the true solution. However, regardless of whether or not we include the fundamental parameter constraints, we are still finding models that match the oscillations constraints reasonably well. Furthermore, recall that the evidence is only the likelihood of obtaining the data, given that the approach is correct.\footnote{In order to obtain correctly normalized probabilities for the different approaches themselves, we have to introduce conditional probabilities like, e.g., $P(\textbf{M2a} | I)$ or $P(\textbf{M2b} | I)$ and use Bayes' theorem. Only comparing the evidence amounts to setting these conditional probabilities to be equal for all tested hypotheses (e.g., $P(\textbf{M2a} | I) = P(\textbf{M2b} | I)$).} We know that the \textbf{a} prior is misrepresenting our state of information. The solar prior approach \textbf{b} more correctly encodes what we know about the Sun, and the age prior \textbf{c} puts even tighter constraints on the pulsation models. Ignoring this information (using prior \textbf{a} and setting equal conditional probabilities) is an interesting and necessary exercise to study the consistency of the results, and how the different models, approaches, and priors work. For an actual detailed study of the solar model physics, however, it is not appropriate. We can nonetheless compare the results, restricting ourselves to the non-adiabatic frequencies and the on average best model for each prior, \textbf{M2}. The resulting parameters are displayed in Table\,\ref{tab:sundist}. 

The results obtained without using our prior knowledge of the Sun for model \textbf{M2a} are spread over several models in the parameter space that can fit the observations quite well. However, for the most probable models, the mass and age are inconsistent with our prior knowledge. These models seem to produce smaller surface effects, and are therefore preferred. For model \textbf{M2b} the situation is similar. Although the mass is now fixed to the true solar value, we do not obtain models that are consistent with the solar age.

For \textbf{M3c} a single combination of physical parameters dominates the results and manages to fit well all the constraints we impose (mass, luminosity, $T_{\rm eff}$, pulsation frequencies, and age). Loosening the conditions on $T_{\rm eff}$ and the luminosity does not significantly change the result. We have also tested slight variations of up to 20 million years in the age prior and do not find the result to be affected. In all cases, we recover a tightly constrained most probable model with $Z_0=0.016$ and $Z_{\rm s} = 0.0144$, and an age of $4.591 \pm 0.005 \rm\,Gyr$. We therefore find a result similar to \cite{houdek2011}. Given that our models take $35 \pm 5\rm\,Myr$ to reach the main sequence, our result is also consistent with meteoritic age determinations of the solar system to within several million years \citep[see, e.g.][]{bouvier2010}. However, we also recover $X_0=0.72$, which leads to an initial helium mass fraction of $Y_0=0.264(1)$. This is different compared to the value of $Y_0 = 0.250(1)$ that was found by \citeauthor{houdek2011}, but more consistent with \cite{asplund2009}. 

{\it Fitting the observations to the adiabatic frequencies, including the age prior, we also recover the exact same model}. We also tested how sensitive the grid is to the prior constraints in order to estimate the actual impact of the pulsation frequencies on the probabilities. If we only evaluate the combined priors, ignoring the frequencies but including the prior on the age, we obtain $X_0=0.71 \pm 0.01$, $Z_0 = 0.019 \pm 0.002$, $Z_{\rm s} = 0.017 \pm 0.002$, $\rm age = 4.603 \pm 0.008 \rm\,Gyr$, and $\alpha_{\rm ml} = 2.1 \pm 0.2$. This leads us to conclude that {\it the frequencies have a decisive impact and actually select the low-metallicity models, no matter whether adiabatic or non-adiabatic model frequencies are used.} 

However, it has to be stressed again that the evidence drops by almost two orders of magnitude when we introduce the age prior. This can be understood by the fact that the solution is so well constrained and at the edge of our current parameter space in $Z_0$, and that many other models can also produce similar pulsation spectra. It could also suggest that we might not have covered the true best model parameters yet in our current grid. Therefore, our next goal will be to extend the grid to lower metallicities, and also include different abundance mixtures, but this is beyond the scope of this paper.  

\begin{figure}[ht!]
   \centering
	   \includegraphics[width=0.45\textwidth]{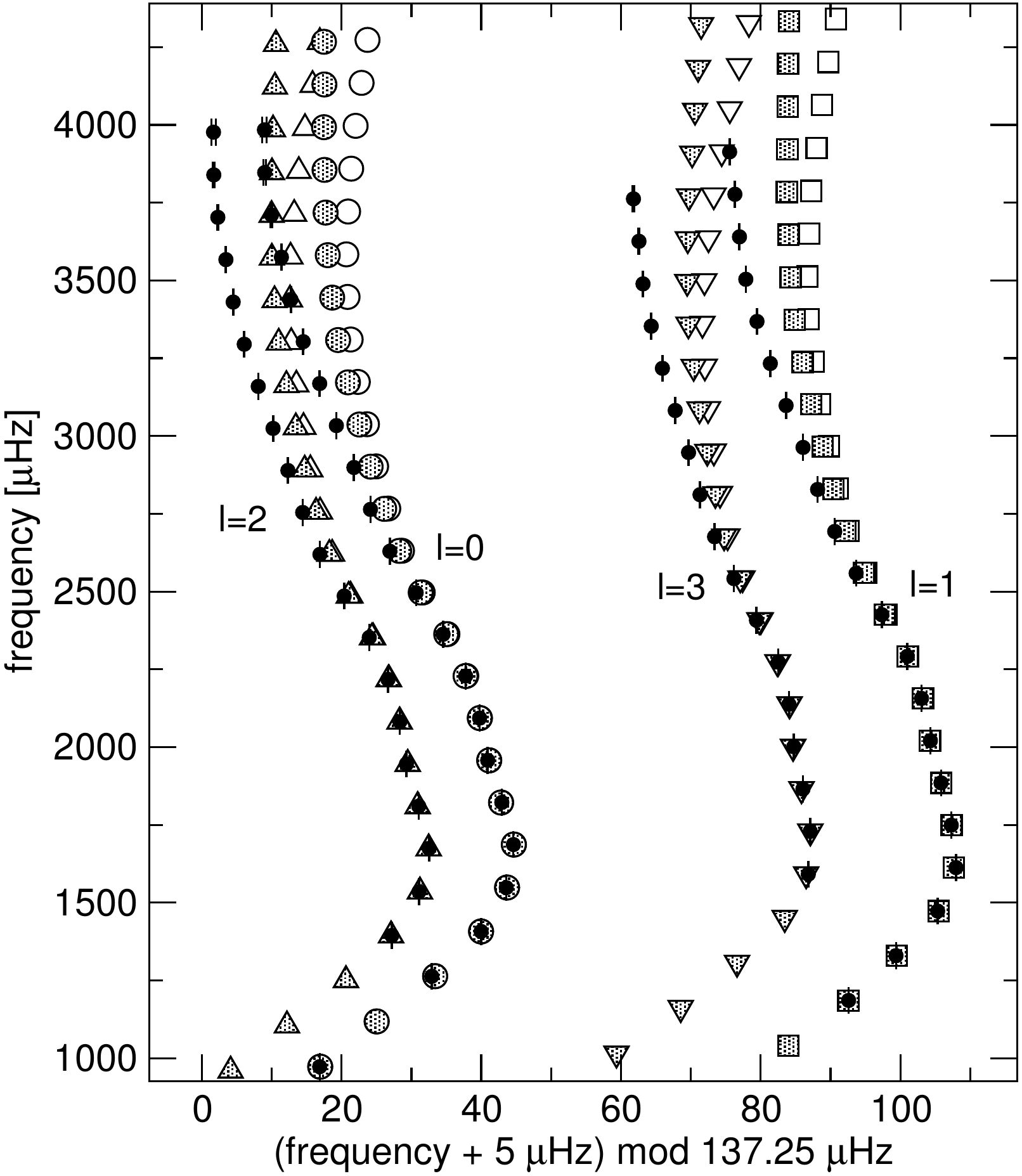}
	   \caption{Non-adiabatic (shaded symbols) and adiabatic (open symbols) frequencies of the most probable solar model from evaluating the BiSON frequencies (black circles + error bars) using approach \textbf{M2c}. Note that frequencies have been shifted upward by 5\uHz, before calculating the $x$-axis values in order to prevent the $l=2$ modes from wrapping around.}
   \label {fig:fit}
\end{figure}

\begin{figure}[ht!]
   \centering
   \includegraphics[width=0.45\textwidth]{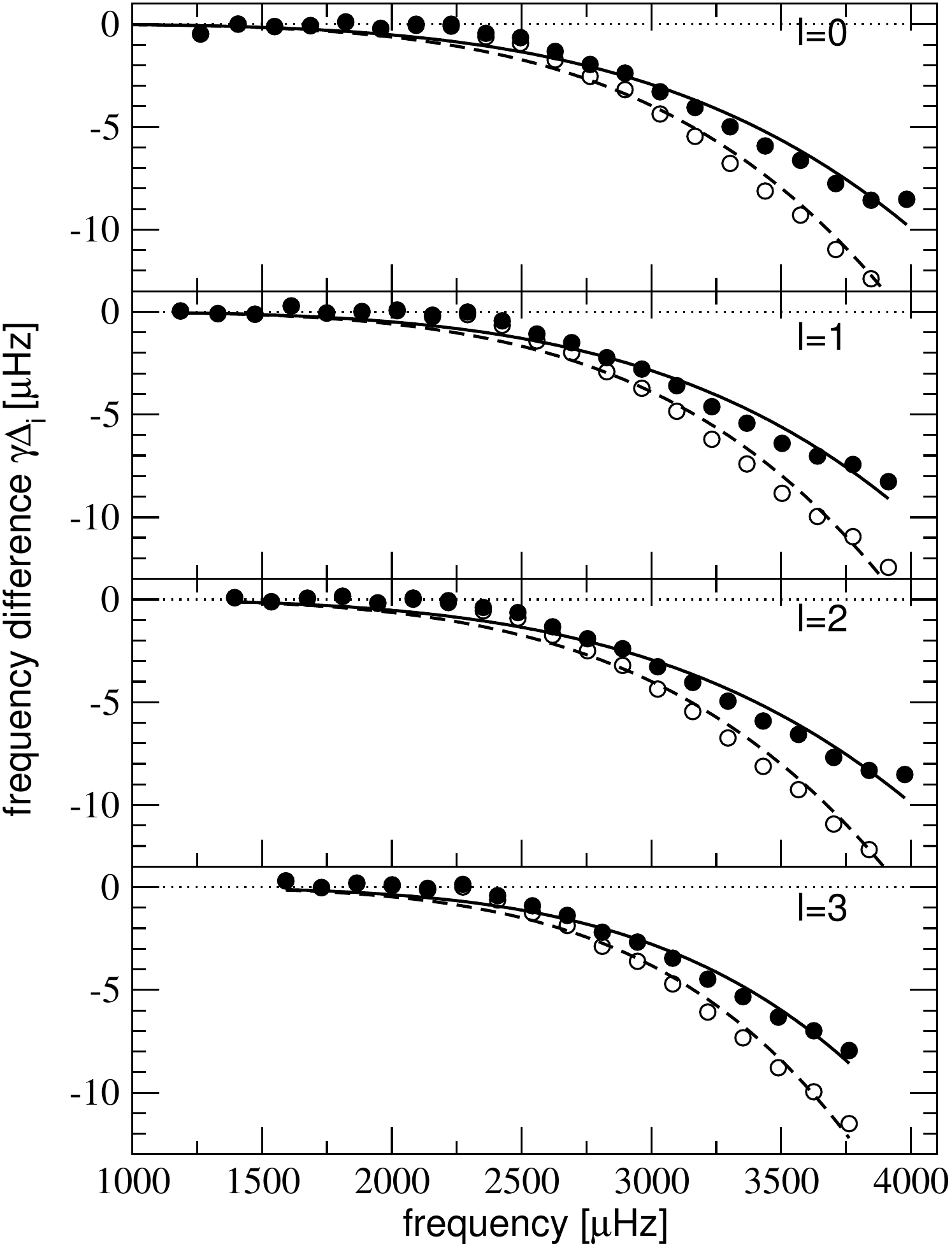}
   \caption{Measured surface effects for non-adiabatic (filled circles) and adiabatic (open circles) frequencies of the most probable solar model from evaluating the BiSON frequencies using approach \textbf{M2c}. The uncertainties of the differences are smaller than the symbols. Least-squares power-law fits (see Equation\,(\ref{equ:kjeldsen})) to the surface effects for the adiabatic (solid line) and non-adiabatic (dashed line) frequencies are also shown.}
   \label {fig:freqdiff_prior}
\end{figure}

Figure\,\ref{fig:fit} compares the BiSON observations with our most probable model at the correct solar age. Even with non-adiabatic frequencies, significant surface effects can still be found. The measured surface effects themselves are shown in Figure\,\ref{fig:freqdiff_prior}, together with least-squares fits following the relation proposed by \cite{kjeldsen2008}. The magnitude of the surface deviations depends on whether the non-adiabatic or the adiabatic frequencies are used for the fit. Nonetheless, our method manages to identify the same exact model to be the most probable, even using the same surface-effect model, thanks to the power of marginalization. However, the non-adiabatic models are vastly preferred in terms of the Bayesian evidence. This is an example for how the presented approach can be used to iterate toward improved stellar model physics, while still recovering meaningful stellar parameters from current asteroseismic investigations. 

We also determined surface-correction power-law exponents for every spherical degree via least-squares fits. For both the non-adiabatic and adiabatic frequencies the best fitting exponents are markedly different from $b=4.9$ which was both advocated by \cite{kjeldsen2008} and also used as the basis for our probabilistic surface model \textbf{M2}. This is also the reason why the \textbf{M3c} models have a greater evidence than their \textbf{M2c} counterparts.
The fitted values range from $b=4.23$ for non-adiabatic ($l=0$) frequencies to $b=5.13$ for adiabatic ($l=3$) frequencies. Moreover, the power-law fits do not match the deviations very well at intermediate radial orders near $2400\,\mu \rm Hz$. From our point of view, fixing the exponent to $b = 4.9$ for a least-squares fit, as for instance done by \cite{brandao2011}, is therefore a potential problem since it does not even match the Sun very well, in particular when improved (e.g., non-adiabatic) physics are implemented. The probabilistic procedure has no problem with these deviations, even though it formally assumes an exponent of $b=4.9$, since the magnitude of the surface effects is marginalized for every frequency.

\subsection{Asteroseismic analysis of a Sun-like star}

To investigate the applicability of our method to current asteroseismic investigations, we also performed an ``asteroseismic" analysis of a Sun-like star, simulated by artificially ``degrading" the set of observed BiSON frequencies to a precision and accuracy expected from current space-based missions for average Sun-like solar-type oscillators. We first multiplied the uncertainties of the BiSON observations by a factor of 20, and then added corresponding random errors to the frequency values. Furthermore, we did not assume to have detailed prior information on the fundamental parameters. Instead, we fitted the ``degraded" data set with a completely flat prior to the same grid as before, again using our surface effect model \textbf{M2}. 

Although a different most probable model is identified, the overall results are comparable to our findings for \textbf{M2a}. They show a slightly larger spread of the model probabilities across the grid. Summarizing the uncertainties for the main parameters by calculating the first and second central moments of the probability distribution in our grid we approximately obtain $M = 1.015 \pm 0.007\, M_{\odot}$, age = $4.76 \pm 0.10\,\rm Gyr$, $X_0 = 0.72 \pm 0.01$, $Z_0 = 0.017 \pm 0.001$, $Z_{\rm s} = 0.0148 \pm 0.0005$, and $\alpha_{\rm ml} = 2.3 \pm 0.1$.

However, these results become worse if we systematically remove lower order modes which are crucial to ``anchor" the surface effect relation. To illustrate, we further degraded our data set by only keeping 13 $l=0$ modes from 1950 to 3580\uHz, 12 $l=1$ modes from 2020 to 3505\uHz, 10 $l=2$ modes between 2080 and 3300\uHz, and 8 $l=3$ modes from 2270 to 3220\uHz. Similar data sets from {\it Kepler} and {\it CoRoT} with comparable uncertainties and numbers of modes have recently been analyzed in the literature. The results for the model parameters become $M = 1.046 \pm 0.007\, M_{\odot}$, age = $4.80 \pm 0.43\,\rm Gyr$, $X_0 = 0.72 \pm 0.01$, $Z_0 = 0.021 \pm 0.01$, $Z_{\rm s} = 0.019 \pm 0.001$, and $\alpha_{\rm ml} = 2.3 \pm 0.1$. Although the values are still within $\sim 5\%$ we are almost at the border of our parameter space, and higher-mass models systematically outperform lower-mass models.

We know from investigating the BiSON data using our grid that we require $\alpha_{\rm ml} = 2.2$ to fit all solar observables. Therefore, in an analysis of a Sun-like star, we can constrain the fit to all models with this value or use a prior based on the marginal posterior probability for $\alpha_{\rm ml}$ as determined from the fit to the Sun. In this case we obtain $M = 1.04 \pm 0.01\, M_{\odot}$, age = $4.41 \pm 0.29\,\rm Gyr$, $X_0 = 0.72 \pm 0.01$, $Z_0 = 0.020 \pm 0.002$, $Z_{\rm s} = 0.018 \pm 0.001$. This is an improvement, but still not comparable to the results obtained when using the full data set.

Thanks to the probabilistic method, however, we can also easily add new observables as further constraints, such as the frequency of maximum power, which can also be inferred from a power spectrum analysis and which approximately scales for Sun-like stars as

\begin{equation}
\nu_{\rm max} \approx \frac{M/M_{\odot}\left(T_{\rm eff}/T_{\rm eff, \odot}\right)^{3.5}}{L/L_{\odot}}\nu_{\rm max, \odot},
\label{equ:nu_max}
\end{equation} 

with $\nu_{\rm max, \odot} = 3120 \pm 5$\uHz\,\citep{kallinger2010c}. Assuming an observed value of $\nu_{\rm max, obs}$ and calculating $\nu_{\rm max, mod}$ for each model according to Equation\,(\ref{equ:nu_max}) we can then multiply the probability for each model with

\begin{equation} 
P(\nu_{\rm max, obs} | M_j^{\Delta}, I) = \frac{1}{\sqrt{2\pi}\sigma_{\nu}}\exp{\left[-\frac{\left(\nu_{\rm max, obs} -\nu_{\rm max, mod}\right)^2}{2\sigma_{\nu}^2}\right]},
\end{equation}
where $\sigma_{\nu} = \sqrt{\sigma^2(\nu_{\rm max, obs}) + \sigma^2(\nu_{\rm max, mod})}$. 

With $\nu_{\rm max, obs} = 3120 \pm 20$\uHz\, for our simulated Sun-like star, we then obtain $M = 1.02 \pm 0.01\, M_{\odot}$, age = $4.39 \pm 0.28\,\rm Gyr$, $X_0 = 0.72 \pm 0.01$, $Z_0 = 0.019 \pm 0.002$, $Z_{\rm s} = 0.017 \pm 0.002$. Finally, if we were able to determine $\nu_{\rm max, obs}$ to about solar precision, the results would be $M = 1.008 \pm 0.006\, M_{\odot}$, age = $4.39 \pm 0.30\,\rm Gyr$, $X_0 = 0.71 \pm 0.01$, $Z_0 = 0.019 \pm 0.002$, $Z_{\rm s} = 0.017 \pm 0.002$. Therefore, if our observations provide precise additional information such as $\nu_{\rm max}$, it can easily be implemented with our method. It then seems possible to obtain reasonably accurate results for Sun-like stars, even in the absence of low-order modes and without a fixed surface effect correction.

\section{Conclusions}

In this paper, we have derived a new, completely probabilistic framework for asteroseismic grid fitting. We explicitly used marginalization and the formulation of combined propositions to allow for the quantitative evaluation of the model grid physics. While computationally more intensive than the standard $\chi^2$ evaluation, this approach has several benefits in that it
\begin{enumerate}
\item allows for the treatment {\it and analysis} of systematic errors such as {\it the surface effects}, therefore removing the need to apply corrections prior to fitting,
\item easily implements uncertainties in the mode identification,
\item takes into account the fact that grids are discrete representations of a continuous parameter space, which is especially important for rapidly varying bumped modes, 
\item provides a consistent framework to use prior knowledge about stellar fundamental parameters or to evaluate additional observables such as $\nu_{\rm max}$, and
\item produces correctly normalized probabilities and likelihoods, respectively evidences, which can be used to assess the model grid physics and the calibration of the grids.
\end{enumerate}
While the above was explicitly derived using the example of a static grid, the probabilistic approach would also be suited for an adaptive grid approach. The Bayesian evidence could be used as a formidable criterion to decide whether an adaptive grid needs to be further refined or not. 

We also showed how to apply our method to study the Sun. The analysis based on our current grid and our prior information matches well the findings of \cite{houdek2011}, and in general fits the up-to-date picture of the Sun. The age of our best model (measured from the pre-main-sequence birth line) is consistent with the meteoritic solar age. The solar model arrives on the ZAMS approximately $35 \pm 5\rm\,Myr$ after appearing on the birth line. {\it We found the same best model whether non-adiabatic or adiabatic frequencies were used.} This shows that our method can adequately deal with different shapes of surface effects, even when using the same (flexible) surface-effect model. One requirement, however, is that there exist enough lower-order modes to ``anchor" the fit.

To our knowledge, this work is also the first completely grid-based asteroseismic analysis of the Sun, using {\it all the information provided by the frequencies} and prior knowledge about the solar fundamental parameters, that results in the need for initial hydrogen, helium and metal mass fractions more consistent with \cite{asplund2009} than the traditional higher-metallicity models. At least for our current grid, these values are required to produce a model that ``looks" like the Sun, pulsates like the Sun, and has the correct solar age. We stress that a formal $\chi^2$ fit to the Sun's oscillation frequencies \citep{guenther2004} or even targeted nonlinear inversion of the oscillation frequencies \citep{marchenkov2000} will not necessarily yield the same model as our approach. With $\chi^2$ fits it is difficult to provide an unbiased correction for surface effects that at the same time does not overly weight the deeper penetrating modes. Some of the deeper penetrating modes are sensitive to the base of the convection zone where the effects of convective overshoot and turbulence, introduced by rotation shears, are not included in the standard models. Inversion methods, where a standard base model is perturbed to fit the oscillations, are also distinct because even though the perturbed model obtained from inversions reveal regions of the standard model that are inadequate, e.g., the base of the convection zone, the inversion model is not an actual standard model in the sense that it is constrained and generated by the model physics.

We know our best-fit model is inaccurate at the surface and we suspect it is inaccurate at the base of the convection zone (the latter suspicion based on the inadequate model physics for this region). Regardless, the model is probabilistically the best model from the current model grid that matches all the known constraints. We speculate that preferring fits that match the oscillation frequencies at the expense of the other physical constraints may be the reason that helioseismologists have been unable reconcile the observed solar {\it p}-mode frequencies with frequencies derived from standard models based on the Asplund mixture and metal abundance \citep{serenelli2009, guzik2010}. We will pursue these matters in a future study where we include model grids based on the Asplund mixture.

While the purpose of our analysis of the Sun is to test the details of our model physics, our method can also be used in general asteroseismic investigations. 
When applying our technique to stars other than the Sun, e.g., recent asteroseismic targets from the {\it Kepler} mission, tight prior constraints as in the solar case are generally not available. However, the probability formalism can simply assign uninformative (e.g., uniform) priors for the unknown parameters and still retain all the remaining benefits like treatment of missing mode identification and of finite grid resolution. 

For current asteroseismology, however, the most important feature is the flexible treatment of the surface effects that differs from the usual approach of employing the empirical correction by \cite{kjeldsen2008} to the frequencies. Instead of measuring the empirical correction for the Sun with the help of a reference model, we use a flexible probabilistic model that allows us to measure surface effects in any star given our current asteroseismic grids. We do not rely on the validity of the solar surface-effect correction and can test new surface-effect models that deviate from the solar power-law approach. Correctly treating the impact of the surface effects on the model probabilities, this also yields correctly propagated uncertainties, and therefore a less biased (but model-dependent) assessment of the stellar fundamental parameters. 

The results presented in the previous section indicate that the accuracy of such current asteroseismic analyses is still an open question and heavily dependent on the number of unaffected, lower-order modes. If there are not enough lower-order modes the surface effect will lead to systematic errors in the fundamental parameter determination. However, even in such a case, by looking at how the evidence changes as better physics are included in the models, our method can be used to iterate toward improved models, hopefully solving the surface-effect problem eventually. 

\acknowledgments{We are very grateful to Werner~W.~Weiss for his valuable input and fruitful discussions. We also thank the referee for improving the quality of the manuscript. MG and DG acknowledge funding from the Natural Sciences \& Engineering Research Council (NSERC) Canada. TK is supported by the FWO-Flanders under project O6260-G.0728.11.}  

\bibliography{bayesfreq}

\begin{thebibliography}{49}
\expandafter\ifx\csname natexlab\endcsname\relax\def\natexlab#1{#1}\fi

\bibitem[{{Alexander} \& {Ferguson}(1994)}]{alexander1994}
{Alexander}, D.~R., \& {Ferguson}, J.~W. 1994, \apj, 437, 879

\bibitem[{{Asplund} {et~al.}(2009){Asplund}, {Grevesse}, {Sauval}, \&
  {Scott}}]{asplund2009}
{Asplund}, M., {Grevesse}, N., {Sauval}, A.~J., \& {Scott}, P. 2009, \araa, 47,
  481

\bibitem[{{Bahcall} {et~al.}(2001){Bahcall}, {Pinsonneault}, \&
  {Basu}}]{bahcall2001}
{Bahcall}, J.~N., {Pinsonneault}, M.~H., \& {Basu}, S. 2001, \apj, 555, 990

\bibitem[{{Bahcall} {et~al.}(1995){Bahcall}, {Pinsonneault}, \&
  {Wasserburg}}]{bahcall1995}
{Bahcall}, J.~N., {Pinsonneault}, M.~H., \& {Wasserburg}, G.~J. 1995, Reviews
  of Modern Physics, 67, 781

\bibitem[{{Bazot} {et~al.}(2008){Bazot}, {Bourguignon}, \&
  {Christensen-Dalsgaard}}]{bazot08}
{Bazot}, M., {Bourguignon}, S., \& {Christensen-Dalsgaard}, J. 2008, \memsai,
  79, 660

\bibitem[{{Bedding} \& {Kjeldsen}(2010)}]{bedding2010a}
{Bedding}, T.~R., \& {Kjeldsen}, H. 2010, Communications in Asteroseismology,
  161, 3

\bibitem[{{Bedding} {et~al.}(2010){Bedding}, {Kjeldsen}, {Campante},
  {Appourchaux}, {Bonanno}, {Chaplin}, {Garcia}, {Marti{\'c}}, {Mosser},
  {Butler}, {Bruntt}, {Kiss}, {O'Toole}, {Kambe}, {Ando}, {Izumiura}, {Sato},
  {Hartmann}, {Hatzes}, {Barban}, {Berthomieu}, {Michel}, {Provost},
  {Turck-Chi{\`e}ze}, {Lebrun}, {Schmitt}, {Bertaux}, {Benatti}, {Claudi},
  {Cosentino}, {Leccia}, {Frandsen}, {Brogaard}, {Glowienka}, {Grundahl},
  {Stempels}, {Arentoft}, {Bazot}, {Christensen-Dalsgaard}, {Dall}, {Karoff},
  {Lundgreen-Nielsen}, {Carrier}, {Eggenberger}, {Sosnowska}, {Wittenmyer},
  {Endl}, {Metcalfe}, {Hekker}, \& {Reffert}}]{bedding2010b}
{Bedding}, T.~R., {et~al.} 2010, \apj, 713, 935

\bibitem[{{Benomar} {et~al.}(2009){Benomar}, {Appourchaux}, \&
  {Baudin}}]{benomar2009}
{Benomar}, O., {Appourchaux}, T., \& {Baudin}, F. 2009, \aap, 506, 15

\bibitem[{{B{\"o}hm-Vitense}(1958)}]{boehm1958}
{B{\"o}hm-Vitense}, E. 1958, \zap, 46, 108

\bibitem[{{Bouvier} \& {Wadhwa}(2010)}]{bouvier2010}
{Bouvier}, A., \& {Wadhwa}, M. 2010, Nature Geoscience, 3, 637

\bibitem[{{Brand{\~a}o} {et~al.}(2011){Brand{\~a}o}, {Do{\u g}an},
  {Christensen-Dalsgaard}, {Cunha}, {Bedding}, {Metcalfe}, {Kjeldsen},
  {Bruntt}, \& {Arentoft}}]{brandao2011}
{Brand{\~a}o}, I.~M., {et~al.} 2011, \aap, 527, A37+

\bibitem[{{Broomhall} {et~al.}(2009){Broomhall}, {Chaplin}, {Davies},
  {Elsworth}, {Fletcher}, {Hale}, {Miller}, \& {New}}]{broomhall2009}
{Broomhall}, A.-M., {Chaplin}, W.~J., {Davies}, G.~R., {Elsworth}, Y.,
  {Fletcher}, S.~T., {Hale}, S.~J., {Miller}, B., \& {New}, R. 2009, \mnras,
  396, L100

\bibitem[{{Chandrasekhar}(1957)}]{chandra1957}
{Chandrasekhar}, S. 1957, {An introduction to the study of stellar structure.},
  ed. {Chandrasekhar, S.}

\bibitem[{{Chaplin} {et~al.}(2010){Chaplin}, {Appourchaux}, {Elsworth},
  {Garc{\'{\i}}a}, {Houdek}, {Karoff}, {Metcalfe}, {Molenda-{\.Z}akowicz},
  {Monteiro}, {Thompson}, {Brown}, {Christensen-Dalsgaard}, {Gilliland},
  {Kjeldsen}, {Borucki}, {Koch}, {Jenkins}, {Ballot}, {Basu}, {Bazot},
  {Bedding}, {Benomar}, {Bonanno}, {Brand{\~a}o}, {Bruntt}, {Campante},
  {Creevey}, {Di Mauro}, {Do{\u g}an}, {Dreizler}, {Eggenberger}, {Esch},
  {Fletcher}, {Frandsen}, {Gai}, {Gaulme}, {Handberg}, {Hekker}, {Howe},
  {Huber}, {Korzennik}, {Lebrun}, {Leccia}, {Martic}, {Mathur}, {Mosser},
  {New}, {Quirion}, {R{\'e}gulo}, {Roxburgh}, {Salabert}, {Schou}, {Sousa},
  {Stello}, {Verner}, {Arentoft}, {Barban}, {Belkacem}, {Benatti}, {Biazzo},
  {Boumier}, {Bradley}, {Broomhall}, {Buzasi}, {Claudi}, {Cunha}, {D'Antona},
  {Deheuvels}, {Derekas}, {Garc{\'{\i}}a Hern{\'a}ndez}, {Giampapa}, {Goupil},
  {Gruberbauer}, {Guzik}, {Hale}, {Ireland}, {Kiss}, {Kitiashvili},
  {Kolenberg}, {Korhonen}, {Kosovichev}, {Kupka}, {Lebreton}, {Leroy},
  {Ludwig}, {Mathis}, {Michel}, {Miglio}, {Montalb{\'a}n}, {Moya}, {Noels},
  {Noyes}, {Pall{\'e}}, {Piau}, {Preston}, {Roca Cort{\'e}s}, {Roth}, {Sato},
  {Schmitt}, {Serenelli}, {Silva Aguirre}, {Stevens}, {Su{\'a}rez}, {Suran},
  {Trampedach}, {Turck-Chi{\`e}ze}, {Uytterhoeven}, {Ventura}, \&
  {Wilson}}]{chaplin2010}
{Chaplin}, W.~J., {et~al.} 2010, \apjl, 713, L169

\bibitem[{{Cohen} \& {Taylor}(1986)}]{cohen1986}
{Cohen}, E.~R., \& {Taylor}, B.~N. 1986, Codata Bulletin No. 63, (New York:
  Pergamon Press)

\bibitem[{{Demarque} {et~al.}(2008){Demarque}, {Guenther}, {Li}, {Mazumdar}, \&
  {Straka}}]{demarque2008}
{Demarque}, P., {Guenther}, D.~B., {Li}, L.~H., {Mazumdar}, A., \& {Straka},
  C.~W. 2008, \apss, 316, 31

\bibitem[{{Deupree} \& {Beslin}(2010)}]{deupree2010}
{Deupree}, R.~G., \& {Beslin}, W. 2010, \apj, 721, 1900

\bibitem[{{Gai} {et~al.}(2011){Gai}, {Basu}, {Chaplin}, \&
  {Elsworth}}]{gai2011}
{Gai}, N., {Basu}, S., {Chaplin}, W.~J., \& {Elsworth}, Y. 2011, \apj, 730, 63

\bibitem[{{Gregory}(2005)}]{gregory2005}
{Gregory}, P.~C. 2005, {Bayesian Logical Data Analysis for the Physical
  Sciences: A Comparative Approach with `Mathematica' Support}, ed. {Gregory,
  P.~C.} (Cambridge University Press)

\bibitem[{{Grevesse} \& {Sauval}(1998)}]{grevesse1998}
{Grevesse}, N., \& {Sauval}, A.~J. 1998, \ssr, 85, 161

\bibitem[{{Gruberbauer} {et~al.}(2009){Gruberbauer}, {Kallinger}, {Weiss}, \&
  {Guenther}}]{gruberbauer09}
{Gruberbauer}, M., {Kallinger}, T., {Weiss}, W.~W., \& {Guenther}, D.~B. 2009,
  \aap, 506, 1043

\bibitem[{{Guenther}(1994)}]{guenther1994}
{Guenther}, D.~B. 1994, \apj, 422, 400

\bibitem[{{Guenther} \& {Brown}(2004)}]{guenther2004}
{Guenther}, D.~B., \& {Brown}, K.~I.~T. 2004, \apj, 600, 419

\bibitem[{{Guenther} {et~al.}(1993){Guenther}, {Pinsonneault}, \&
  {Bahcall}}]{guenther1993}
{Guenther}, D.~B., {Pinsonneault}, M.~H., \& {Bahcall}, J.~N. 1993, \apj, 418,
  469

\bibitem[{{Guzik} \& {Mussack}(2010)}]{guzik2010}
{Guzik}, J.~A., \& {Mussack}, K. 2010, \apj, 713, 1108

\bibitem[{{Handberg} \& {Campante}(2011)}]{handberg11}
{Handberg}, R., \& {Campante}, T.~L. 2011, \aap, 527, A56+

\bibitem[{{Hayashi}(1961)}]{hayashi1961}
{Hayashi}, C. 1961, \pasj, 13, 450

\bibitem[{{Hickey} \& {Alton}(1983)}]{hickey1983}
{Hickey}, J.~R., \& {Alton}, B.~M. 1983, in Solar Irradiance Variations of
  Active Region Time Scales, NASA Conference Publication 2310, ed. B.~.J.
  LaBonte, G.~A. Chapman, H.~S. Hudson, \& R.~C. Wilson, 43

\bibitem[{{Houdek} \& {Gough}(2011)}]{houdek2011}
{Houdek}, G., \& {Gough}, D.~O. 2011, \mnras, 418, 1217

\bibitem[{{Huber} {et~al.}(2011){Huber}, {Bedding}, {Stello}, {Hekker},
  {Mathur}, {Mosser}, {Verner}, {Bonanno}, {Buzasi}, {Campante}, {Elsworth},
  {Hale}, {Kallinger}, {Silva Aguirre}, {Chaplin}, {De Ridder},
  {Garc{\'{\i}}a}, {Appourchaux}, {Frandsen}, {Houdek}, {Molenda-{\.Z}akowicz},
  {Monteiro}, {Christensen-Dalsgaard}, {Gilliland}, {Kawaler}, {Kjeldsen},
  {Broomhall}, {Corsaro}, {Salabert}, {Sanderfer}, {Seader}, \&
  {Smith}}]{huber2011}
{Huber}, D., {et~al.} 2011, \apj, 743, 143

\bibitem[{{Iglesias} \& {Rogers}(1996)}]{iglesias1996}
{Iglesias}, C.~A., \& {Rogers}, F.~J. 1996, \apj, 464, 943

\bibitem[{{Jaynes} \& {Bretthorst}(2003)}]{jaynes}
{Jaynes}, E.~T., \& {Bretthorst}, G.~L. 2003, {Probability Theory}, ed.
  {Jaynes, E.~T.~\& Bretthorst, G.~L.} (Cambridge University Press)

\bibitem[{{J{\o}rgensen} \& {Lindegren}(2005)}]{jorgensen2005}
{J{\o}rgensen}, B.~R., \& {Lindegren}, L. 2005, \aap, 436, 127

\bibitem[{{Kallinger} {et~al.}(2010{\natexlab{a}}){Kallinger}, {Gruberbauer},
  {Guenther}, {Fossati}, \& {Weiss}}]{kallinger2010b}
{Kallinger}, T., {Gruberbauer}, M., {Guenther}, D.~B., {Fossati}, L., \&
  {Weiss}, W.~W. 2010{\natexlab{a}}, \aap, 510, A106+

\bibitem[{{Kallinger} {et~al.}(2010{\natexlab{b}}){Kallinger}, {Mosser},
  {Hekker}, {Huber}, {Stello}, {Mathur}, {Basu}, {Bedding}, {Chaplin}, {De
  Ridder}, {Elsworth}, {Frandsen}, {Garc{\'{\i}}a}, {Gruberbauer}, {Matthews},
  {Borucki}, {Bruntt}, {Christensen-Dalsgaard}, {Gilliland}, {Kjeldsen}, \&
  {Koch}}]{kallinger2010c}
{Kallinger}, T., {et~al.} 2010{\natexlab{b}}, \aap, 522, A1

\bibitem[{{Kallinger} {et~al.}(2010{\natexlab{c}}){Kallinger}, {Weiss},
  {Barban}, {Baudin}, {Cameron}, {Carrier}, {De Ridder}, {Goupil},
  {Gruberbauer}, {Hatzes}, {Hekker}, {Samadi}, \& {Deleuil}}]{kallinger2010a}
---. 2010{\natexlab{c}}, \aap, 509, A77+

\bibitem[{{Kjeldsen} {et~al.}(2008){Kjeldsen}, {Bedding}, \&
  {Christensen-Dalsgaard}}]{kjeldsen2008}
{Kjeldsen}, H., {Bedding}, T.~R., \& {Christensen-Dalsgaard}, J. 2008, \apjl,
  683, L175

\bibitem[{{Krishna Swamy}(1966)}]{krishna1966}
{Krishna Swamy}, K.~S. 1966, \apj, 145, 174

\bibitem[{{Lane}(1869)}]{lane1869}
{Lane}, J.~H. 1869, Amer. J. Sci., 2nd ser., 50, 57

\bibitem[{{Marchenkov} {et~al.}(2000){Marchenkov}, {Roxburgh}, \&
  {Vorontsov}}]{marchenkov2000}
{Marchenkov}, K., {Roxburgh}, I., \& {Vorontsov}, S. 2000, \mnras, 312, 39

\bibitem[{{Mathur} {et~al.}(2010){Mathur}, {Garc{\'{\i}}a}, {R{\'e}gulo},
  {Creevey}, {Ballot}, {Salabert}, {Arentoft}, {Quirion}, {Chaplin}, \&
  {Kjeldsen}}]{mathur2010}
{Mathur}, S., {et~al.} 2010, \aap, 511, A46+

\bibitem[{{Metcalfe} {et~al.}(2010){Metcalfe}, {Monteiro}, {Thompson},
  {Molenda-{\.Z}akowicz}, {Appourchaux}, {Chaplin}, {Do{\u g}an},
  {Eggenberger}, {Bedding}, {Bruntt}, {Creevey}, {Quirion}, {Stello},
  {Bonanno}, {Silva Aguirre}, {Basu}, {Esch}, {Gai}, {Di Mauro}, {Kosovichev},
  {Kitiashvili}, {Su{\'a}rez}, {Moya}, {Piau}, {Garc{\'{\i}}a}, {Marques},
  {Frasca}, {Biazzo}, {Sousa}, {Dreizler}, {Bazot}, {Karoff}, {Frandsen},
  {Wilson}, {Brown}, {Christensen-Dalsgaard}, {Gilliland}, {Kjeldsen},
  {Campante}, {Fletcher}, {Handberg}, {R{\'e}gulo}, {Salabert}, {Schou},
  {Verner}, {Ballot}, {Broomhall}, {Elsworth}, {Hekker}, {Huber}, {Mathur},
  {New}, {Roxburgh}, {Sato}, {White}, {Borucki}, {Koch}, \&
  {Jenkins}}]{metcalfe2010}
{Metcalfe}, T.~S., {et~al.} 2010, \apj, 723, 1583

\bibitem[{{Pont} \& {Eyer}(2004)}]{pont2004}
{Pont}, F., \& {Eyer}, L. 2004, \mnras, 351, 487

\bibitem[{{Quirion} {et~al.}(2010){Quirion}, {Christensen-Dalsgaard}, \&
  {Arentoft}}]{quirion2010}
{Quirion}, P.-O., {Christensen-Dalsgaard}, J., \& {Arentoft}, T. 2010, \apj,
  725, 2176

\bibitem[{{Rogers}(1986)}]{rogers1986}
{Rogers}, F.~J. 1986, \apj, 310, 723

\bibitem[{{Rogers} {et~al.}(1996){Rogers}, {Swenson}, \&
  {Iglesias}}]{rogers1996}
{Rogers}, F.~J., {Swenson}, F.~J., \& {Iglesias}, C.~A. 1996, \apj, 456, 902

\bibitem[{{Roxburgh}(2005)}]{roxburgh2005}
{Roxburgh}, I.~W. 2005, \aap, 434, 665

\bibitem[{{Serenelli} {et~al.}(2009){Serenelli}, {Basu}, {Ferguson}, \&
  {Asplund}}]{serenelli2009}
{Serenelli}, A.~M., {Basu}, S., {Ferguson}, J.~W., \& {Asplund}, M. 2009,
  \apjl, 705, L123

\bibitem[{{Takeda} {et~al.}(2007){Takeda}, {Ford}, {Sills}, {Rasio}, {Fischer},
  \& {Valenti}}]{takeda2007}
{Takeda}, G., {Ford}, E.~B., {Sills}, A., {Rasio}, F.~A., {Fischer}, D.~A., \&
  {Valenti}, J.~A. 2007, \apjs, 168, 297

\end{thebibliography}

 \end{document}